%%%%%%%%%%%%%%%%%%%%%%%%%% mic44.tex %%%%%%%%%%%%%%%%%%%%%%%%%%%%%%%%
%==============================================================================
% PSDUMP.TEX
% TeX macros for dumping included Postscript to files.
% Adapted from Knuth's \answer macro in the TeXbook.
% Requires Plain TeX.  Maybe other flavors will work too?
% Jamie Stephens, jamies@math.utexas.edu, 28 Nov 94

\def\endofps{EndOfTheIncludedPostscriptMagicCookie}
\chardef\other=12
\newwrite\psdumphandle
\outer\def\psdump#1{\par\medbreak
  \immediate\openout\psdumphandle=#1
  \copytoblankline}
\def\copytoblankline{\begingroup\setupcopy\copypsline}
\def\setupcopy{\def\do##1{\catcode`##1=\other}\dospecials
  \catcode`\\=\other \obeylines}
{\obeylines \gdef\copypsline#1
  {\def\next{#1}%
  \ifx\next\endofps\let\next=\endgroup %
  \else\immediate\write\psdumphandle{\next} \let\next=\copypsline\fi\next}}
\outer\def\closepsdump{
  \immediate\closeout\psdumphandle}

%==============================================================================
\psdump{nmicf1.ps}%!
/st {stroke} def
/m {moveto} def
/rm {rmoveto} def
/l {lineto} def
/rl {rlineto} def
0.5 setlinewidth 
209.764 56.693 m
243.780 56.693 l  
243.780 283.465 l  
209.764 283.465 l  
209.764 56.693 l  
gsave closepath 0.9 setgray fill grestore 
st
0.5 setlinewidth 
56.693 170.079 m
396.850 170.079 l  
-4.252 2.835 rl  
4.252 -2.835 rl  
-4.252 -2.835 rl  
4.252 2.835 rl  
st
226.772 0.000 m
226.772 325.984 l  
-2.835 -4.252 rl  
2.835 4.252 rl  
2.835 -4.252 rl  
-2.835 4.252 rl  
st
113.386 170.079 m
226.772 56.693 l  
340.157 170.079 l  
226.772 283.465 l  
113.386 170.079 l  
st
1 setlinewidth 
226.772 56.693 m
226.028 57.827 l  
225.289 58.961 l  
224.554 60.094 l  
223.824 61.228 l  
223.098 62.362 l  
222.377 63.496 l  
221.660 64.630 l  
220.947 65.764 l  
220.240 66.898 l  
219.537 68.031 l  
218.839 69.165 l  
218.146 70.299 l  
217.458 71.433 l  
216.775 72.567 l  
216.097 73.701 l  
215.424 74.835 l  
214.756 75.969 l  
214.093 77.102 l  
213.436 78.236 l  
212.784 79.370 l  
212.138 80.504 l  
211.497 81.638 l  
210.862 82.772 l  
210.232 83.906 l  
209.609 85.039 l  
208.991 86.173 l  
208.378 87.307 l  
207.772 88.441 l  
207.172 89.575 l  
206.578 90.709 l  
205.990 91.843 l  
205.408 92.976 l  
204.832 94.110 l  
204.263 95.244 l  
203.701 96.378 l  
203.144 97.512 l  
202.595 98.646 l  
202.052 99.780 l  
201.516 100.913 l  
200.986 102.047 l  
200.464 103.181 l  
199.948 104.315 l  
199.439 105.449 l  
198.938 106.583 l  
198.444 107.717 l  
197.957 108.850 l  
197.477 109.984 l  
197.005 111.118 l  
196.540 112.252 l  
196.083 113.386 l  
195.633 114.520 l  
195.191 115.654 l  
194.757 116.787 l  
194.330 117.921 l  
193.912 119.055 l  
193.502 120.189 l  
193.099 121.323 l  
192.705 122.457 l  
192.319 123.591 l  
191.941 124.724 l  
191.571 125.858 l  
191.210 126.992 l  
190.858 128.126 l  
190.513 129.260 l  
190.178 130.394 l  
189.851 131.528 l  
189.533 132.661 l  
189.223 133.795 l  
188.923 134.929 l  
188.631 136.063 l  
188.348 137.197 l  
188.075 138.331 l  
187.810 139.465 l  
187.555 140.598 l  
187.308 141.732 l  
187.071 142.866 l  
186.843 144.000 l  
186.625 145.134 l  
186.416 146.268 l  
186.216 147.402 l  
186.026 148.535 l  
185.845 149.669 l  
185.674 150.803 l  
185.512 151.937 l  
185.360 153.071 l  
185.218 154.205 l  
185.085 155.339 l  
184.963 156.472 l  
184.849 157.606 l  
184.746 158.740 l  
184.652 159.874 l  
184.568 161.008 l  
184.494 162.142 l  
184.430 163.276 l  
184.376 164.409 l  
184.331 165.543 l  
184.296 166.677 l  
184.272 167.811 l  
184.257 168.945 l  
184.252 170.079 l  
184.257 171.213 l  
184.272 172.346 l  
184.296 173.480 l  
184.331 174.614 l  
184.376 175.748 l  
184.430 176.882 l  
184.494 178.016 l  
184.568 179.150 l  
184.652 180.283 l  
184.746 181.417 l  
184.849 182.551 l  
184.963 183.685 l  
185.085 184.819 l  
185.218 185.953 l  
185.360 187.087 l  
185.512 188.220 l  
185.674 189.354 l  
185.845 190.488 l  
186.026 191.622 l  
186.216 192.756 l  
186.416 193.890 l  
186.625 195.024 l  
186.843 196.157 l  
187.071 197.291 l  
187.308 198.425 l  
187.555 199.559 l  
187.810 200.693 l  
188.075 201.827 l  
188.348 202.961 l  
188.631 204.094 l  
188.923 205.228 l  
189.223 206.362 l  
189.533 207.496 l  
189.851 208.630 l  
190.178 209.764 l  
190.513 210.898 l  
190.858 212.031 l  
191.210 213.165 l  
191.571 214.299 l  
191.941 215.433 l  
192.319 216.567 l  
192.705 217.701 l  
193.099 218.835 l  
193.502 219.969 l  
193.912 221.102 l  
194.330 222.236 l  
194.757 223.370 l  
195.191 224.504 l  
195.633 225.638 l  
196.083 226.772 l  
196.540 227.906 l  
197.005 229.039 l  
197.477 230.173 l  
197.957 231.307 l  
198.444 232.441 l  
198.938 233.575 l  
199.439 234.709 l  
199.948 235.843 l  
200.464 236.976 l  
200.986 238.110 l  
201.516 239.244 l  
202.052 240.378 l  
202.595 241.512 l  
203.144 242.646 l  
203.701 243.780 l  
204.263 244.913 l  
204.832 246.047 l  
205.408 247.181 l  
205.990 248.315 l  
206.578 249.449 l  
207.172 250.583 l  
207.772 251.717 l  
208.378 252.850 l  
208.991 253.984 l  
209.609 255.118 l  
210.232 256.252 l  
210.862 257.386 l  
211.497 258.520 l  
212.138 259.654 l  
212.784 260.787 l  
213.436 261.921 l  
214.093 263.055 l  
214.756 264.189 l  
215.424 265.323 l  
216.097 266.457 l  
216.775 267.591 l  
217.458 268.724 l  
218.146 269.858 l  
218.839 270.992 l  
219.537 272.126 l  
220.240 273.260 l  
220.947 274.394 l  
221.660 275.528 l  
222.377 276.661 l  
223.098 277.795 l  
223.824 278.929 l  
224.554 280.063 l  
225.289 281.197 l  
226.028 282.331 l  
226.772 283.465 l  
226.772 56.693 m
226.603 57.827 l  
226.437 58.961 l  
226.272 60.094 l  
226.108 61.228 l  
225.947 62.362 l  
225.787 63.496 l  
225.628 64.630 l  
225.472 65.764 l  
225.317 66.898 l  
225.163 68.031 l  
225.012 69.165 l  
224.861 70.299 l  
224.713 71.433 l  
224.566 72.567 l  
224.421 73.701 l  
224.278 74.835 l  
224.136 75.969 l  
223.996 77.102 l  
223.858 78.236 l  
223.721 79.370 l  
223.586 80.504 l  
223.453 81.638 l  
223.321 82.772 l  
223.191 83.906 l  
223.063 85.039 l  
222.936 86.173 l  
222.811 87.307 l  
222.688 88.441 l  
222.567 89.575 l  
222.447 90.709 l  
222.328 91.843 l  
222.212 92.976 l  
222.097 94.110 l  
221.984 95.244 l  
221.872 96.378 l  
221.763 97.512 l  
221.654 98.646 l  
221.548 99.780 l  
221.443 100.913 l  
221.340 102.047 l  
221.239 103.181 l  
221.139 104.315 l  
221.041 105.449 l  
220.945 106.583 l  
220.850 107.717 l  
220.757 108.850 l  
220.666 109.984 l  
220.577 111.118 l  
220.489 112.252 l  
220.403 113.386 l  
220.318 114.520 l  
220.236 115.654 l  
220.154 116.787 l  
220.075 117.921 l  
219.998 119.055 l  
219.922 120.189 l  
219.847 121.323 l  
219.775 122.457 l  
219.704 123.591 l  
219.635 124.724 l  
219.567 125.858 l  
219.502 126.992 l  
219.438 128.126 l  
219.375 129.260 l  
219.315 130.394 l  
219.256 131.528 l  
219.198 132.661 l  
219.143 133.795 l  
219.089 134.929 l  
219.037 136.063 l  
218.987 137.197 l  
218.938 138.331 l  
218.891 139.465 l  
218.846 140.598 l  
218.802 141.732 l  
218.760 142.866 l  
218.720 144.000 l  
218.682 145.134 l  
218.645 146.268 l  
218.610 147.402 l  
218.576 148.535 l  
218.545 149.669 l  
218.515 150.803 l  
218.487 151.937 l  
218.460 153.071 l  
218.435 154.205 l  
218.412 155.339 l  
218.391 156.472 l  
218.371 157.606 l  
218.353 158.740 l  
218.337 159.874 l  
218.322 161.008 l  
218.310 162.142 l  
218.299 163.276 l  
218.289 164.409 l  
218.281 165.543 l  
218.275 166.677 l  
218.271 167.811 l  
218.269 168.945 l  
218.268 170.079 l  
218.269 171.213 l  
218.271 172.346 l  
218.275 173.480 l  
218.281 174.614 l  
218.289 175.748 l  
218.299 176.882 l  
218.310 178.016 l  
218.322 179.150 l  
218.337 180.283 l  
218.353 181.417 l  
218.371 182.551 l  
218.391 183.685 l  
218.412 184.819 l  
218.435 185.953 l  
218.460 187.087 l  
218.487 188.220 l  
218.515 189.354 l  
218.545 190.488 l  
218.576 191.622 l  
218.610 192.756 l  
218.645 193.890 l  
218.682 195.024 l  
218.720 196.157 l  
218.760 197.291 l  
218.802 198.425 l  
218.846 199.559 l  
218.891 200.693 l  
218.938 201.827 l  
218.987 202.961 l  
219.037 204.094 l  
219.089 205.228 l  
219.143 206.362 l  
219.198 207.496 l  
219.256 208.630 l  
219.315 209.764 l  
219.375 210.898 l  
219.438 212.031 l  
219.502 213.165 l  
219.567 214.299 l  
219.635 215.433 l  
219.704 216.567 l  
219.775 217.701 l  
219.847 218.835 l  
219.922 219.969 l  
219.998 221.102 l  
220.075 222.236 l  
220.154 223.370 l  
220.236 224.504 l  
220.318 225.638 l  
220.403 226.772 l  
220.489 227.906 l  
220.577 229.039 l  
220.666 230.173 l  
220.757 231.307 l  
220.850 232.441 l  
220.945 233.575 l  
221.041 234.709 l  
221.139 235.843 l  
221.239 236.976 l  
221.340 238.110 l  
221.443 239.244 l  
221.548 240.378 l  
221.654 241.512 l  
221.763 242.646 l  
221.872 243.780 l  
221.984 244.913 l  
222.097 246.047 l  
222.212 247.181 l  
222.328 248.315 l  
222.447 249.449 l  
222.567 250.583 l  
222.688 251.717 l  
222.811 252.850 l  
222.936 253.984 l  
223.063 255.118 l  
223.191 256.252 l  
223.321 257.386 l  
223.453 258.520 l  
223.586 259.654 l  
223.721 260.787 l  
223.858 261.921 l  
223.996 263.055 l  
224.136 264.189 l  
224.278 265.323 l  
224.421 266.457 l  
224.566 267.591 l  
224.713 268.724 l  
224.861 269.858 l  
225.012 270.992 l  
225.163 272.126 l  
225.317 273.260 l  
225.472 274.394 l  
225.628 275.528 l  
225.787 276.661 l  
225.947 277.795 l  
226.108 278.929 l  
226.272 280.063 l  
226.437 281.197 l  
226.603 282.331 l  
226.772 283.465 l  
226.772 56.693 m
226.940 57.827 l  
227.107 58.961 l  
227.272 60.094 l  
227.435 61.228 l  
227.597 62.362 l  
227.757 63.496 l  
227.915 64.630 l  
228.072 65.764 l  
228.227 66.898 l  
228.380 68.031 l  
228.532 69.165 l  
228.682 70.299 l  
228.830 71.433 l  
228.977 72.567 l  
229.122 73.701 l  
229.265 74.835 l  
229.407 75.969 l  
229.547 77.102 l  
229.685 78.236 l  
229.822 79.370 l  
229.957 80.504 l  
230.090 81.638 l  
230.222 82.772 l  
230.352 83.906 l  
230.480 85.039 l  
230.607 86.173 l  
230.732 87.307 l  
230.855 88.441 l  
230.977 89.575 l  
231.097 90.709 l  
231.215 91.843 l  
231.331 92.976 l  
231.446 94.110 l  
231.560 95.244 l  
231.671 96.378 l  
231.781 97.512 l  
231.889 98.646 l  
231.995 99.780 l  
232.100 100.913 l  
232.203 102.047 l  
232.304 103.181 l  
232.404 104.315 l  
232.502 105.449 l  
232.598 106.583 l  
232.693 107.717 l  
232.786 108.850 l  
232.877 109.984 l  
232.967 111.118 l  
233.054 112.252 l  
233.141 113.386 l  
233.225 114.520 l  
233.308 115.654 l  
233.389 116.787 l  
233.468 117.921 l  
233.546 119.055 l  
233.622 120.189 l  
233.696 121.323 l  
233.769 122.457 l  
233.839 123.591 l  
233.909 124.724 l  
233.976 125.858 l  
234.042 126.992 l  
234.106 128.126 l  
234.168 129.260 l  
234.229 130.394 l  
234.288 131.528 l  
234.345 132.661 l  
234.400 133.795 l  
234.454 134.929 l  
234.506 136.063 l  
234.557 137.197 l  
234.605 138.331 l  
234.652 139.465 l  
234.698 140.598 l  
234.741 141.732 l  
234.783 142.866 l  
234.823 144.000 l  
234.862 145.134 l  
234.899 146.268 l  
234.934 147.402 l  
234.967 148.535 l  
234.999 149.669 l  
235.028 150.803 l  
235.057 151.937 l  
235.083 153.071 l  
235.108 154.205 l  
235.131 155.339 l  
235.152 156.472 l  
235.172 157.606 l  
235.190 158.740 l  
235.206 159.874 l  
235.221 161.008 l  
235.234 162.142 l  
235.245 163.276 l  
235.254 164.409 l  
235.262 165.543 l  
235.268 166.677 l  
235.272 167.811 l  
235.275 168.945 l  
235.276 170.079 l  
235.275 171.213 l  
235.272 172.346 l  
235.268 173.480 l  
235.262 174.614 l  
235.254 175.748 l  
235.245 176.882 l  
235.234 178.016 l  
235.221 179.150 l  
235.206 180.283 l  
235.190 181.417 l  
235.172 182.551 l  
235.152 183.685 l  
235.131 184.819 l  
235.108 185.953 l  
235.083 187.087 l  
235.057 188.220 l  
235.028 189.354 l  
234.999 190.488 l  
234.967 191.622 l  
234.934 192.756 l  
234.899 193.890 l  
234.862 195.024 l  
234.823 196.157 l  
234.783 197.291 l  
234.741 198.425 l  
234.698 199.559 l  
234.652 200.693 l  
234.605 201.827 l  
234.557 202.961 l  
234.506 204.094 l  
234.454 205.228 l  
234.400 206.362 l  
234.345 207.496 l  
234.288 208.630 l  
234.229 209.764 l  
234.168 210.898 l  
234.106 212.031 l  
234.042 213.165 l  
233.976 214.299 l  
233.909 215.433 l  
233.839 216.567 l  
233.769 217.701 l  
233.696 218.835 l  
233.622 219.969 l  
233.546 221.102 l  
233.468 222.236 l  
233.389 223.370 l  
233.308 224.504 l  
233.225 225.638 l  
233.141 226.772 l  
233.054 227.906 l  
232.967 229.039 l  
232.877 230.173 l  
232.786 231.307 l  
232.693 232.441 l  
232.598 233.575 l  
232.502 234.709 l  
232.404 235.843 l  
232.304 236.976 l  
232.203 238.110 l  
232.100 239.244 l  
231.995 240.378 l  
231.889 241.512 l  
231.781 242.646 l  
231.671 243.780 l  
231.560 244.913 l  
231.446 246.047 l  
231.331 247.181 l  
231.215 248.315 l  
231.097 249.449 l  
230.977 250.583 l  
230.855 251.717 l  
230.732 252.850 l  
230.607 253.984 l  
230.480 255.118 l  
230.352 256.252 l  
230.222 257.386 l  
230.090 258.520 l  
229.957 259.654 l  
229.822 260.787 l  
229.685 261.921 l  
229.547 263.055 l  
229.407 264.189 l  
229.265 265.323 l  
229.122 266.457 l  
228.977 267.591 l  
228.830 268.724 l  
228.682 269.858 l  
228.532 270.992 l  
228.380 272.126 l  
228.227 273.260 l  
228.072 274.394 l  
227.915 275.528 l  
227.757 276.661 l  
227.597 277.795 l  
227.435 278.929 l  
227.272 280.063 l  
227.107 281.197 l  
226.940 282.331 l  
226.772 283.465 l  
226.772 56.693 m
227.303 57.827 l  
227.830 58.961 l  
228.354 60.094 l  
228.872 61.228 l  
229.387 62.362 l  
229.897 63.496 l  
230.403 64.630 l  
230.905 65.764 l  
231.402 66.898 l  
231.895 68.031 l  
232.384 69.165 l  
232.868 70.299 l  
233.347 71.433 l  
233.822 72.567 l  
234.292 73.701 l  
234.758 74.835 l  
235.220 75.969 l  
235.676 77.102 l  
236.128 78.236 l  
236.576 79.370 l  
237.018 80.504 l  
237.456 81.638 l  
237.889 82.772 l  
238.318 83.906 l  
238.741 85.039 l  
239.160 86.173 l  
239.574 87.307 l  
239.982 88.441 l  
240.386 89.575 l  
240.785 90.709 l  
241.180 91.843 l  
241.569 92.976 l  
241.953 94.110 l  
242.332 95.244 l  
242.706 96.378 l  
243.074 97.512 l  
243.438 98.646 l  
243.797 99.780 l  
244.150 100.913 l  
244.498 102.047 l  
244.841 103.181 l  
245.179 104.315 l  
245.511 105.449 l  
245.839 106.583 l  
246.160 107.717 l  
246.477 108.850 l  
246.788 109.984 l  
247.094 111.118 l  
247.394 112.252 l  
247.689 113.386 l  
247.978 114.520 l  
248.262 115.654 l  
248.541 116.787 l  
248.814 117.921 l  
249.081 119.055 l  
249.343 120.189 l  
249.599 121.323 l  
249.850 122.457 l  
250.095 123.591 l  
250.334 124.724 l  
250.568 125.858 l  
250.796 126.992 l  
251.018 128.126 l  
251.235 129.260 l  
251.446 130.394 l  
251.651 131.528 l  
251.850 132.661 l  
252.044 133.795 l  
252.232 134.929 l  
252.414 136.063 l  
252.590 137.197 l  
252.761 138.331 l  
252.925 139.465 l  
253.084 140.598 l  
253.237 141.732 l  
253.384 142.866 l  
253.525 144.000 l  
253.660 145.134 l  
253.789 146.268 l  
253.912 147.402 l  
254.029 148.535 l  
254.141 149.669 l  
254.246 150.803 l  
254.345 151.937 l  
254.439 153.071 l  
254.526 154.205 l  
254.608 155.339 l  
254.683 156.472 l  
254.753 157.606 l  
254.816 158.740 l  
254.873 159.874 l  
254.925 161.008 l  
254.970 162.142 l  
255.009 163.276 l  
255.043 164.409 l  
255.070 165.543 l  
255.091 166.677 l  
255.106 167.811 l  
255.115 168.945 l  
255.118 170.079 l  
255.115 171.213 l  
255.106 172.346 l  
255.091 173.480 l  
255.070 174.614 l  
255.043 175.748 l  
255.009 176.882 l  
254.970 178.016 l  
254.925 179.150 l  
254.873 180.283 l  
254.816 181.417 l  
254.753 182.551 l  
254.683 183.685 l  
254.608 184.819 l  
254.526 185.953 l  
254.439 187.087 l  
254.345 188.220 l  
254.246 189.354 l  
254.141 190.488 l  
254.029 191.622 l  
253.912 192.756 l  
253.789 193.890 l  
253.660 195.024 l  
253.525 196.157 l  
253.384 197.291 l  
253.237 198.425 l  
253.084 199.559 l  
252.925 200.693 l  
252.761 201.827 l  
252.590 202.961 l  
252.414 204.094 l  
252.232 205.228 l  
252.044 206.362 l  
251.850 207.496 l  
251.651 208.630 l  
251.446 209.764 l  
251.235 210.898 l  
251.018 212.031 l  
250.796 213.165 l  
250.568 214.299 l  
250.334 215.433 l  
250.095 216.567 l  
249.850 217.701 l  
249.599 218.835 l  
249.343 219.969 l  
249.081 221.102 l  
248.814 222.236 l  
248.541 223.370 l  
248.262 224.504 l  
247.978 225.638 l  
247.689 226.772 l  
247.394 227.906 l  
247.094 229.039 l  
246.788 230.173 l  
246.477 231.307 l  
246.160 232.441 l  
245.839 233.575 l  
245.511 234.709 l  
245.179 235.843 l  
244.841 236.976 l  
244.498 238.110 l  
244.150 239.244 l  
243.797 240.378 l  
243.438 241.512 l  
243.074 242.646 l  
242.706 243.780 l  
242.332 244.913 l  
241.953 246.047 l  
241.569 247.181 l  
241.180 248.315 l  
240.785 249.449 l  
240.386 250.583 l  
239.982 251.717 l  
239.574 252.850 l  
239.160 253.984 l  
238.741 255.118 l  
238.318 256.252 l  
237.889 257.386 l  
237.456 258.520 l  
237.018 259.654 l  
236.576 260.787 l  
236.128 261.921 l  
235.676 263.055 l  
235.220 264.189 l  
234.758 265.323 l  
234.292 266.457 l  
233.822 267.591 l  
233.347 268.724 l  
232.868 269.858 l  
232.384 270.992 l  
231.895 272.126 l  
231.402 273.260 l  
230.905 274.394 l  
230.403 275.528 l  
229.897 276.661 l  
229.387 277.795 l  
228.872 278.929 l  
228.354 280.063 l  
227.830 281.197 l  
227.303 282.331 l  
226.772 283.465 l  
st
0.5 setlinewidth 
239.528 279.213 m
0.000 2.835 rl  
0.000 -2.835 rl  
2.835 0.000 rl  
-2.835 0.000 rl  
267.874 307.559 l  
st
234.558 226.772 m
0.000 2.835 rl  
0.000 -2.835 rl  
2.835 0.000 rl  
-2.835 0.000 rl  
277.078 269.291 l  
st
226.772 127.559 m
2.835 1.417 rl  
-2.835 -1.417 rl  
2.835 -2.835 rl  
-2.835 2.835 rl  
311.811 113.386 l  
st
200.905 99.213 m
-2.835 0.000 rl  
2.835 0.000 rl  
0.000 -2.835 rl  
0.000 2.835 rl  
178.228 76.535 l  
st
141.732 155.906 m
-2.835 0.000 rl  
2.835 0.000 rl  
0.000 -2.835 rl  
0.000 2.835 rl  
113.386 127.559 l  
st
st
showpage 

EndOfTheIncludedPostscriptMagicCookie
\closepsdump
\psdump{nmicf2.ps}%!
/st {stroke} def
/m {moveto} def
/rm {rmoveto} def
/l {lineto} def
/rl {rlineto} def
0.5 setlinewidth 
0.5 setlinewidth 
0.000 141.732 m
198.425 141.732 l  
st
170.079 141.732 m
170.070 142.845 l  
170.044 143.958 l  
170.000 145.071 l  
169.939 146.182 l  
169.860 147.292 l  
169.764 148.401 l  
169.651 149.509 l  
169.520 150.614 l  
169.372 151.717 l  
169.206 152.818 l  
169.023 153.916 l  
168.824 155.011 l  
168.606 156.103 l  
168.372 157.191 l  
168.121 158.276 l  
167.852 159.356 l  
167.567 160.432 l  
167.265 161.503 l  
166.946 162.570 l  
166.610 163.631 l  
166.258 164.687 l  
165.889 165.737 l  
165.504 166.782 l  
165.102 167.820 l  
164.684 168.852 l  
164.250 169.877 l  
163.800 170.895 l  
163.334 171.906 l  
162.852 172.909 l  
162.355 173.905 l  
161.842 174.893 l  
161.313 175.872 l  
160.769 176.844 l  
160.210 177.806 l  
159.636 178.760 l  
159.047 179.704 l  
158.443 180.639 l  
157.825 181.565 l  
157.192 182.481 l  
156.545 183.386 l  
155.883 184.282 l  
155.208 185.167 l  
154.519 186.041 l  
153.816 186.904 l  
153.100 187.756 l  
152.370 188.597 l  
151.627 189.426 l  
150.872 190.243 l  
150.103 191.049 l  
149.323 191.842 l  
148.529 192.623 l  
147.724 193.391 l  
146.906 194.147 l  
146.077 194.890 l  
145.236 195.619 l  
144.384 196.336 l  
143.521 197.038 l  
142.647 197.728 l  
141.762 198.403 l  
140.867 199.064 l  
139.961 199.711 l  
139.045 200.344 l  
138.120 200.963 l  
137.185 201.567 l  
136.240 202.156 l  
135.286 202.730 l  
134.324 203.289 l  
133.353 203.833 l  
132.373 204.361 l  
131.385 204.874 l  
130.389 205.372 l  
129.386 205.854 l  
128.375 206.320 l  
127.357 206.770 l  
126.332 207.204 l  
125.300 207.622 l  
124.262 208.024 l  
123.218 208.409 l  
122.167 208.778 l  
121.111 209.130 l  
120.050 209.466 l  
118.984 209.785 l  
117.912 210.087 l  
116.836 210.372 l  
115.756 210.640 l  
114.672 210.892 l  
113.583 211.126 l  
112.492 211.343 l  
111.397 211.543 l  
110.299 211.726 l  
109.198 211.891 l  
108.094 212.040 l  
106.989 212.170 l  
105.882 212.284 l  
104.773 212.380 l  
103.662 212.459 l  
102.551 212.520 l  
101.439 212.563 l  
100.326 212.590 l  
99.213 212.598 l  
98.099 212.590 l  
96.987 212.563 l  
95.874 212.520 l  
94.763 212.459 l  
93.653 212.380 l  
92.544 212.284 l  
91.436 212.170 l  
90.331 212.040 l  
89.227 211.891 l  
88.127 211.726 l  
87.029 211.543 l  
85.934 211.343 l  
84.842 211.126 l  
83.754 210.892 l  
82.669 210.640 l  
81.589 210.372 l  
80.513 210.087 l  
79.442 209.785 l  
78.375 209.466 l  
77.314 209.130 l  
76.258 208.778 l  
75.208 208.409 l  
74.163 208.024 l  
73.125 207.622 l  
72.093 207.204 l  
71.068 206.770 l  
70.050 206.320 l  
69.039 205.854 l  
68.036 205.372 l  
67.040 204.874 l  
66.052 204.361 l  
65.073 203.833 l  
64.101 203.289 l  
63.139 202.730 l  
62.185 202.156 l  
61.241 201.567 l  
60.305 200.963 l  
59.380 200.344 l  
58.464 199.711 l  
57.559 199.064 l  
56.663 198.403 l  
55.778 197.728 l  
54.904 197.038 l  
54.041 196.336 l  
53.189 195.619 l  
52.348 194.890 l  
51.519 194.147 l  
50.701 193.391 l  
49.896 192.623 l  
49.103 191.842 l  
48.322 191.049 l  
47.553 190.243 l  
46.798 189.426 l  
46.055 188.597 l  
45.326 187.756 l  
44.609 186.904 l  
43.907 186.041 l  
43.217 185.167 l  
42.542 184.282 l  
41.881 183.386 l  
41.233 182.481 l  
40.601 181.565 l  
39.982 180.639 l  
39.378 179.704 l  
38.789 178.760 l  
38.215 177.806 l  
37.656 176.844 l  
37.112 175.872 l  
36.584 174.893 l  
36.070 173.905 l  
35.573 172.909 l  
35.091 171.906 l  
34.625 170.895 l  
34.175 169.877 l  
33.741 168.852 l  
33.323 167.820 l  
32.921 166.782 l  
32.536 165.737 l  
32.167 164.687 l  
31.815 163.631 l  
31.479 162.570 l  
31.160 161.503 l  
30.858 160.432 l  
30.573 159.356 l  
30.304 158.276 l  
30.053 157.191 l  
29.819 156.103 l  
29.602 155.011 l  
29.402 153.916 l  
29.219 152.818 l  
29.053 151.717 l  
28.905 150.614 l  
28.774 149.509 l  
28.661 148.401 l  
28.565 147.292 l  
28.486 146.182 l  
28.425 145.071 l  
28.381 143.958 l  
28.355 142.845 l  
28.346 141.732 l  
28.355 140.619 l  
28.381 139.506 l  
28.425 138.394 l  
28.486 137.283 l  
28.565 136.172 l  
28.661 135.063 l  
28.774 133.956 l  
28.905 132.850 l  
29.053 131.747 l  
29.219 130.646 l  
29.402 129.548 l  
29.602 128.453 l  
29.819 127.362 l  
30.053 126.273 l  
30.304 125.189 l  
30.573 124.109 l  
30.858 123.033 l  
31.160 121.961 l  
31.479 120.895 l  
31.815 119.833 l  
32.167 118.778 l  
32.536 117.727 l  
32.921 116.683 l  
33.323 115.645 l  
33.741 114.613 l  
34.175 113.588 l  
34.625 112.570 l  
35.091 111.559 l  
35.573 110.555 l  
36.070 109.560 l  
36.584 108.572 l  
37.112 107.592 l  
37.656 106.621 l  
38.215 105.658 l  
38.789 104.705 l  
39.378 103.760 l  
39.982 102.825 l  
40.601 101.900 l  
41.233 100.984 l  
41.881 100.078 l  
42.542 99.183 l  
43.217 98.298 l  
43.907 97.424 l  
44.609 96.561 l  
45.326 95.708 l  
46.055 94.868 l  
46.798 94.038 l  
47.553 93.221 l  
48.322 92.416 l  
49.103 91.622 l  
49.896 90.841 l  
50.701 90.073 l  
51.519 89.317 l  
52.348 88.575 l  
53.189 87.845 l  
54.041 87.129 l  
54.904 86.426 l  
55.778 85.737 l  
56.663 85.062 l  
57.559 84.400 l  
58.464 83.753 l  
59.380 83.120 l  
60.305 82.502 l  
61.241 81.898 l  
62.185 81.309 l  
63.139 80.735 l  
64.101 80.176 l  
65.073 79.632 l  
66.052 79.103 l  
67.040 78.590 l  
68.036 78.093 l  
69.039 77.611 l  
70.050 77.145 l  
71.068 76.695 l  
72.093 76.261 l  
73.125 75.843 l  
74.163 75.441 l  
75.208 75.056 l  
76.258 74.687 l  
77.314 74.335 l  
78.375 73.999 l  
79.442 73.680 l  
80.513 73.378 l  
81.589 73.093 l  
82.669 72.824 l  
83.754 72.573 l  
84.842 72.339 l  
85.934 72.121 l  
87.029 71.921 l  
88.127 71.739 l  
89.227 71.573 l  
90.331 71.425 l  
91.436 71.294 l  
92.544 71.181 l  
93.653 71.085 l  
94.763 71.006 l  
95.874 70.945 l  
96.987 70.901 l  
98.099 70.875 l  
99.213 70.866 l  
100.326 70.875 l  
101.439 70.901 l  
102.551 70.945 l  
103.662 71.006 l  
104.773 71.085 l  
105.882 71.181 l  
106.989 71.294 l  
108.094 71.425 l  
109.198 71.573 l  
110.299 71.739 l  
111.397 71.921 l  
112.492 72.121 l  
113.583 72.339 l  
114.672 72.573 l  
115.756 72.824 l  
116.836 73.093 l  
117.912 73.378 l  
118.984 73.680 l  
120.050 73.999 l  
121.111 74.335 l  
122.167 74.687 l  
123.218 75.056 l  
124.262 75.441 l  
125.300 75.843 l  
126.332 76.261 l  
127.357 76.695 l  
128.375 77.145 l  
129.386 77.611 l  
130.389 78.093 l  
131.385 78.590 l  
132.373 79.103 l  
133.353 79.632 l  
134.324 80.176 l  
135.286 80.735 l  
136.240 81.309 l  
137.185 81.898 l  
138.120 82.502 l  
139.045 83.120 l  
139.961 83.753 l  
140.867 84.400 l  
141.762 85.062 l  
142.647 85.737 l  
143.521 86.426 l  
144.384 87.129 l  
145.236 87.845 l  
146.077 88.575 l  
146.906 89.317 l  
147.724 90.073 l  
148.529 90.841 l  
149.323 91.622 l  
150.103 92.416 l  
150.872 93.221 l  
151.627 94.038 l  
152.370 94.868 l  
153.100 95.708 l  
153.816 96.561 l  
154.519 97.424 l  
155.208 98.298 l  
155.883 99.183 l  
156.545 100.078 l  
157.192 100.984 l  
157.825 101.900 l  
158.443 102.825 l  
159.047 103.760 l  
159.636 104.705 l  
160.210 105.658 l  
160.769 106.621 l  
161.313 107.592 l  
161.842 108.572 l  
162.355 109.560 l  
162.852 110.555 l  
163.334 111.559 l  
163.800 112.570 l  
164.250 113.588 l  
164.684 114.613 l  
165.102 115.645 l  
165.504 116.683 l  
165.889 117.727 l  
166.258 118.778 l  
166.610 119.833 l  
166.946 120.895 l  
167.265 121.961 l  
167.567 123.033 l  
167.852 124.109 l  
168.121 125.189 l  
168.372 126.273 l  
168.606 127.362 l  
168.824 128.453 l  
169.023 129.548 l  
169.206 130.646 l  
169.372 131.747 l  
169.520 132.850 l  
169.651 133.956 l  
169.764 135.063 l  
169.860 136.172 l  
169.939 137.283 l  
170.000 138.394 l  
170.044 139.506 l  
170.070 140.619 l  
170.079 141.732 l  
149.323 191.842 m
5.669 0.000 rl  
-5.669 0.000 rl  
0.000 -5.669 rl  
0.000 5.669 rl  
st
155.906 141.732 m
155.899 141.866 l  
155.878 141.999 l  
155.843 142.133 l  
155.794 142.266 l  
155.731 142.399 l  
155.654 142.533 l  
155.563 142.665 l  
155.458 142.798 l  
155.340 142.930 l  
155.208 143.063 l  
155.061 143.194 l  
154.901 143.326 l  
154.728 143.457 l  
154.540 143.587 l  
154.339 143.717 l  
154.124 143.847 l  
153.896 143.976 l  
153.654 144.105 l  
153.399 144.233 l  
153.131 144.360 l  
152.849 144.487 l  
152.554 144.613 l  
152.246 144.738 l  
151.924 144.863 l  
151.590 144.987 l  
151.243 145.110 l  
150.883 145.232 l  
150.510 145.353 l  
150.124 145.473 l  
149.726 145.593 l  
149.316 145.712 l  
148.893 145.829 l  
148.458 145.946 l  
148.011 146.061 l  
147.551 146.176 l  
147.080 146.289 l  
146.597 146.401 l  
146.102 146.512 l  
145.596 146.622 l  
145.078 146.731 l  
144.549 146.838 l  
144.009 146.944 l  
143.457 147.049 l  
142.895 147.153 l  
142.322 147.255 l  
141.739 147.356 l  
141.144 147.456 l  
140.540 147.554 l  
139.925 147.650 l  
139.301 147.745 l  
138.666 147.839 l  
138.022 147.931 l  
137.368 148.022 l  
136.704 148.111 l  
136.032 148.199 l  
135.350 148.285 l  
134.659 148.369 l  
133.960 148.452 l  
133.252 148.533 l  
132.536 148.612 l  
131.811 148.690 l  
131.079 148.766 l  
130.338 148.840 l  
129.590 148.912 l  
128.835 148.983 l  
128.072 149.052 l  
127.302 149.119 l  
126.525 149.184 l  
125.741 149.248 l  
124.951 149.309 l  
124.154 149.369 l  
123.351 149.427 l  
122.543 149.483 l  
121.728 149.537 l  
120.908 149.589 l  
120.083 149.639 l  
119.252 149.687 l  
118.417 149.733 l  
117.576 149.778 l  
116.732 149.820 l  
115.883 149.860 l  
115.029 149.899 l  
114.172 149.935 l  
113.312 149.969 l  
112.447 150.001 l  
111.580 150.031 l  
110.709 150.060 l  
109.836 150.086 l  
108.960 150.110 l  
108.081 150.132 l  
107.201 150.151 l  
106.318 150.169 l  
105.434 150.185 l  
104.548 150.198 l  
103.661 150.210 l  
102.772 150.219 l  
101.883 150.227 l  
100.993 150.232 l  
100.103 150.235 l  
99.213 150.236 l  
98.322 150.235 l  
97.432 150.232 l  
96.542 150.227 l  
95.653 150.219 l  
94.765 150.210 l  
93.877 150.198 l  
92.991 150.185 l  
92.107 150.169 l  
91.224 150.151 l  
90.344 150.132 l  
89.465 150.110 l  
88.589 150.086 l  
87.716 150.060 l  
86.845 150.031 l  
85.978 150.001 l  
85.114 149.969 l  
84.253 149.935 l  
83.396 149.899 l  
82.543 149.860 l  
81.694 149.820 l  
80.849 149.778 l  
80.009 149.733 l  
79.173 149.687 l  
78.343 149.639 l  
77.517 149.589 l  
76.697 149.537 l  
75.883 149.483 l  
75.074 149.427 l  
74.271 149.369 l  
73.475 149.309 l  
72.684 149.248 l  
71.901 149.184 l  
71.124 149.119 l  
70.354 149.052 l  
69.591 148.983 l  
68.835 148.912 l  
68.087 148.840 l  
67.346 148.766 l  
66.614 148.690 l  
65.889 148.612 l  
65.173 148.533 l  
64.465 148.452 l  
63.766 148.369 l  
63.075 148.285 l  
62.393 148.199 l  
61.721 148.111 l  
61.058 148.022 l  
60.404 147.931 l  
59.759 147.839 l  
59.125 147.745 l  
58.500 147.650 l  
57.885 147.554 l  
57.281 147.456 l  
56.687 147.356 l  
56.103 147.255 l  
55.530 147.153 l  
54.968 147.049 l  
54.416 146.944 l  
53.876 146.838 l  
53.347 146.731 l  
52.829 146.622 l  
52.323 146.512 l  
51.828 146.401 l  
51.345 146.289 l  
50.874 146.176 l  
50.415 146.061 l  
49.967 145.946 l  
49.532 145.829 l  
49.109 145.712 l  
48.699 145.593 l  
48.301 145.473 l  
47.915 145.353 l  
47.542 145.232 l  
47.182 145.110 l  
46.835 144.987 l  
46.501 144.863 l  
46.180 144.738 l  
45.871 144.613 l  
45.576 144.487 l  
45.294 144.360 l  
45.026 144.233 l  
44.771 144.105 l  
44.529 143.976 l  
44.301 143.847 l  
44.086 143.717 l  
43.885 143.587 l  
43.698 143.457 l  
43.524 143.326 l  
43.364 143.194 l  
43.218 143.063 l  
43.085 142.930 l  
42.967 142.798 l  
42.862 142.665 l  
42.771 142.533 l  
42.694 142.399 l  
42.632 142.266 l  
42.583 142.133 l  
42.548 141.999 l  
42.527 141.866 l  
42.520 141.732 l  
42.527 141.599 l  
42.548 141.465 l  
42.583 141.332 l  
42.632 141.198 l  
42.694 141.065 l  
42.771 140.932 l  
42.862 140.799 l  
42.967 140.666 l  
43.085 140.534 l  
43.218 140.402 l  
43.364 140.270 l  
43.524 140.139 l  
43.698 140.008 l  
43.885 139.877 l  
44.086 139.747 l  
44.301 139.617 l  
44.529 139.488 l  
44.771 139.360 l  
45.026 139.232 l  
45.294 139.104 l  
45.576 138.978 l  
45.871 138.852 l  
46.180 138.726 l  
46.501 138.602 l  
46.835 138.478 l  
47.182 138.355 l  
47.542 138.233 l  
47.915 138.111 l  
48.301 137.991 l  
48.699 137.872 l  
49.109 137.753 l  
49.532 137.635 l  
49.967 137.519 l  
50.415 137.403 l  
50.874 137.289 l  
51.345 137.176 l  
51.828 137.063 l  
52.323 136.952 l  
52.829 136.842 l  
53.347 136.734 l  
53.876 136.626 l  
54.416 136.520 l  
54.968 136.415 l  
55.530 136.312 l  
56.103 136.209 l  
56.687 136.109 l  
57.281 136.009 l  
57.885 135.911 l  
58.500 135.814 l  
59.125 135.719 l  
59.759 135.625 l  
60.404 135.533 l  
61.058 135.443 l  
61.721 135.353 l  
62.393 135.266 l  
63.075 135.180 l  
63.766 135.096 l  
64.465 135.013 l  
65.173 134.932 l  
65.889 134.852 l  
66.614 134.775 l  
67.346 134.699 l  
68.087 134.625 l  
68.835 134.552 l  
69.591 134.481 l  
70.354 134.413 l  
71.124 134.345 l  
71.901 134.280 l  
72.684 134.217 l  
73.475 134.155 l  
74.271 134.096 l  
75.074 134.038 l  
75.883 133.982 l  
76.697 133.928 l  
77.517 133.876 l  
78.343 133.826 l  
79.173 133.777 l  
80.009 133.731 l  
80.849 133.687 l  
81.694 133.645 l  
82.543 133.604 l  
83.396 133.566 l  
84.253 133.530 l  
85.114 133.496 l  
85.978 133.463 l  
86.845 133.433 l  
87.716 133.405 l  
88.589 133.379 l  
89.465 133.355 l  
90.344 133.333 l  
91.224 133.313 l  
92.107 133.295 l  
92.991 133.280 l  
93.877 133.266 l  
94.765 133.255 l  
95.653 133.245 l  
96.542 133.238 l  
97.432 133.233 l  
98.322 133.229 l  
99.213 133.228 l  
100.103 133.229 l  
100.993 133.233 l  
101.883 133.238 l  
102.772 133.245 l  
103.661 133.255 l  
104.548 133.266 l  
105.434 133.280 l  
106.318 133.295 l  
107.201 133.313 l  
108.081 133.333 l  
108.960 133.355 l  
109.836 133.379 l  
110.709 133.405 l  
111.580 133.433 l  
112.447 133.463 l  
113.312 133.496 l  
114.172 133.530 l  
115.029 133.566 l  
115.883 133.604 l  
116.732 133.645 l  
117.576 133.687 l  
118.417 133.731 l  
119.252 133.777 l  
120.083 133.826 l  
120.908 133.876 l  
121.728 133.928 l  
122.543 133.982 l  
123.351 134.038 l  
124.154 134.096 l  
124.951 134.155 l  
125.741 134.217 l  
126.525 134.280 l  
127.302 134.345 l  
128.072 134.413 l  
128.835 134.481 l  
129.590 134.552 l  
130.338 134.625 l  
131.079 134.699 l  
131.811 134.775 l  
132.536 134.852 l  
133.252 134.932 l  
133.960 135.013 l  
134.659 135.096 l  
135.350 135.180 l  
136.032 135.266 l  
136.704 135.353 l  
137.368 135.443 l  
138.022 135.533 l  
138.666 135.625 l  
139.301 135.719 l  
139.925 135.814 l  
140.540 135.911 l  
141.144 136.009 l  
141.739 136.109 l  
142.322 136.209 l  
142.895 136.312 l  
143.457 136.415 l  
144.009 136.520 l  
144.549 136.626 l  
145.078 136.734 l  
145.596 136.842 l  
146.102 136.952 l  
146.597 137.063 l  
147.080 137.176 l  
147.551 137.289 l  
148.011 137.403 l  
148.458 137.519 l  
148.893 137.635 l  
149.316 137.753 l  
149.726 137.872 l  
150.124 137.991 l  
150.510 138.111 l  
150.883 138.233 l  
151.243 138.355 l  
151.590 138.478 l  
151.924 138.602 l  
152.246 138.726 l  
152.554 138.852 l  
152.849 138.978 l  
153.131 139.104 l  
153.399 139.232 l  
153.654 139.360 l  
153.896 139.488 l  
154.124 139.617 l  
154.339 139.747 l  
154.540 139.877 l  
154.728 140.008 l  
154.901 140.139 l  
155.061 140.270 l  
155.208 140.402 l  
155.340 140.534 l  
155.458 140.666 l  
155.563 140.799 l  
155.654 140.932 l  
155.731 141.065 l  
155.794 141.198 l  
155.843 141.332 l  
155.878 141.465 l  
155.899 141.599 l  
155.906 141.732 l  
closepath 0.8 setgray fill 0 setgray 
0.000 149.097 m
0.283 149.072 l  
0.567 149.047 l  
0.850 149.022 l  
1.134 148.996 l  
1.417 148.970 l  
1.701 148.944 l  
1.984 148.917 l  
2.268 148.890 l  
2.551 148.862 l  
2.835 148.834 l  
3.118 148.806 l  
3.402 148.778 l  
3.685 148.749 l  
3.969 148.719 l  
4.252 148.690 l  
4.535 148.660 l  
4.819 148.629 l  
5.102 148.598 l  
5.386 148.567 l  
5.669 148.535 l  
5.953 148.503 l  
6.236 148.471 l  
6.520 148.438 l  
6.803 148.404 l  
7.087 148.371 l  
7.370 148.336 l  
7.654 148.302 l  
7.937 148.266 l  
8.220 148.231 l  
8.504 148.195 l  
8.787 148.158 l  
9.071 148.121 l  
9.354 148.083 l  
9.638 148.045 l  
9.921 148.007 l  
10.205 147.967 l  
10.488 147.928 l  
10.772 147.888 l  
11.055 147.847 l  
11.339 147.805 l  
11.622 147.763 l  
11.906 147.721 l  
12.189 147.678 l  
12.472 147.634 l  
12.756 147.589 l  
13.039 147.544 l  
13.323 147.499 l  
13.606 147.452 l  
13.890 147.405 l  
14.173 147.357 l  
14.457 147.309 l  
14.740 147.259 l  
15.024 147.209 l  
15.307 147.158 l  
15.591 147.106 l  
15.874 147.054 l  
16.157 147.000 l  
16.441 146.946 l  
16.724 146.891 l  
17.008 146.835 l  
17.291 146.777 l  
17.575 146.719 l  
17.858 146.660 l  
18.142 146.600 l  
18.425 146.538 l  
18.709 146.475 l  
18.992 146.412 l  
19.276 146.346 l  
19.559 146.280 l  
19.843 146.212 l  
20.126 146.143 l  
20.409 146.072 l  
20.693 145.999 l  
20.976 145.925 l  
21.260 145.849 l  
21.543 145.771 l  
21.827 145.692 l  
22.110 145.610 l  
22.394 145.526 l  
22.677 145.439 l  
22.961 145.350 l  
23.244 145.258 l  
23.528 145.163 l  
23.811 145.065 l  
24.094 144.963 l  
24.378 144.858 l  
24.661 144.748 l  
24.945 144.634 l  
25.228 144.514 l  
25.512 144.388 l  
25.795 144.255 l  
26.079 144.113 l  
26.362 143.962 l  
26.646 143.800 l  
26.929 143.622 l  
27.213 143.425 l  
27.496 143.200 l  
27.780 142.932 l  
28.063 142.582 l  
28.346 141.732 l  
28.063 140.883 l  
27.780 140.533 l  
27.496 140.265 l  
27.213 140.040 l  
26.929 139.843 l  
26.646 139.665 l  
26.362 139.502 l  
26.079 139.351 l  
25.795 139.210 l  
25.512 139.077 l  
25.228 138.951 l  
24.945 138.831 l  
24.661 138.716 l  
24.378 138.607 l  
24.094 138.501 l  
23.811 138.399 l  
23.528 138.301 l  
23.244 138.206 l  
22.961 138.115 l  
22.677 138.026 l  
22.394 137.939 l  
22.110 137.855 l  
21.827 137.773 l  
21.543 137.693 l  
21.260 137.615 l  
20.976 137.539 l  
20.693 137.465 l  
20.409 137.393 l  
20.126 137.322 l  
19.843 137.253 l  
19.559 137.185 l  
19.276 137.118 l  
18.992 137.053 l  
18.709 136.989 l  
18.425 136.926 l  
18.142 136.865 l  
17.858 136.805 l  
17.575 136.745 l  
17.291 136.687 l  
17.008 136.630 l  
16.724 136.574 l  
16.441 136.518 l  
16.157 136.464 l  
15.874 136.411 l  
15.591 136.358 l  
15.307 136.306 l  
15.024 136.255 l  
14.740 136.205 l  
14.457 136.156 l  
14.173 136.107 l  
13.890 136.060 l  
13.606 136.012 l  
13.323 135.966 l  
13.039 135.920 l  
12.756 135.875 l  
12.472 135.831 l  
12.189 135.787 l  
11.906 135.744 l  
11.622 135.701 l  
11.339 135.659 l  
11.055 135.618 l  
10.772 135.577 l  
10.488 135.537 l  
10.205 135.497 l  
9.921 135.458 l  
9.638 135.419 l  
9.354 135.381 l  
9.071 135.344 l  
8.787 135.306 l  
8.504 135.270 l  
8.220 135.234 l  
7.937 135.198 l  
7.654 135.163 l  
7.370 135.128 l  
7.087 135.094 l  
6.803 135.060 l  
6.520 135.027 l  
6.236 134.994 l  
5.953 134.961 l  
5.669 134.929 l  
5.386 134.897 l  
5.102 134.866 l  
4.819 134.835 l  
4.535 134.805 l  
4.252 134.775 l  
3.969 134.745 l  
3.685 134.716 l  
3.402 134.687 l  
3.118 134.658 l  
2.835 134.630 l  
2.551 134.602 l  
2.268 134.575 l  
1.984 134.548 l  
1.701 134.521 l  
1.417 134.495 l  
1.134 134.469 l  
0.850 134.443 l  
0.567 134.417 l  
0.283 134.392 l  
0.000 134.368 l  
closepath 0.8 setgray fill 0 setgray 
198.425 149.097 m
198.142 149.072 l  
197.858 149.047 l  
197.575 149.022 l  
197.291 148.996 l  
197.008 148.970 l  
196.724 148.944 l  
196.441 148.917 l  
196.157 148.890 l  
195.874 148.862 l  
195.591 148.834 l  
195.307 148.806 l  
195.024 148.778 l  
194.740 148.749 l  
194.457 148.719 l  
194.173 148.690 l  
193.890 148.660 l  
193.606 148.629 l  
193.323 148.598 l  
193.039 148.567 l  
192.756 148.535 l  
192.472 148.503 l  
192.189 148.471 l  
191.906 148.438 l  
191.622 148.404 l  
191.339 148.371 l  
191.055 148.336 l  
190.772 148.302 l  
190.488 148.266 l  
190.205 148.231 l  
189.921 148.195 l  
189.638 148.158 l  
189.354 148.121 l  
189.071 148.083 l  
188.787 148.045 l  
188.504 148.007 l  
188.220 147.967 l  
187.937 147.928 l  
187.654 147.888 l  
187.370 147.847 l  
187.087 147.805 l  
186.803 147.763 l  
186.520 147.721 l  
186.236 147.678 l  
185.953 147.634 l  
185.669 147.589 l  
185.386 147.544 l  
185.102 147.499 l  
184.819 147.452 l  
184.535 147.405 l  
184.252 147.357 l  
183.969 147.309 l  
183.685 147.259 l  
183.402 147.209 l  
183.118 147.158 l  
182.835 147.106 l  
182.551 147.054 l  
182.268 147.000 l  
181.984 146.946 l  
181.701 146.891 l  
181.417 146.835 l  
181.134 146.777 l  
180.850 146.719 l  
180.567 146.660 l  
180.283 146.600 l  
180.000 146.538 l  
179.717 146.475 l  
179.433 146.412 l  
179.150 146.346 l  
178.866 146.280 l  
178.583 146.212 l  
178.299 146.143 l  
178.016 146.072 l  
177.732 145.999 l  
177.449 145.925 l  
177.165 145.849 l  
176.882 145.771 l  
176.598 145.692 l  
176.315 145.610 l  
176.031 145.526 l  
175.748 145.439 l  
175.465 145.350 l  
175.181 145.258 l  
174.898 145.163 l  
174.614 145.065 l  
174.331 144.963 l  
174.047 144.858 l  
173.764 144.748 l  
173.480 144.634 l  
173.197 144.514 l  
172.913 144.388 l  
172.630 144.255 l  
172.346 144.113 l  
172.063 143.962 l  
171.780 143.800 l  
171.496 143.622 l  
171.213 143.425 l  
170.929 143.200 l  
170.646 142.932 l  
170.362 142.582 l  
170.079 141.732 l  
170.362 140.883 l  
170.646 140.533 l  
170.929 140.265 l  
171.213 140.040 l  
171.496 139.843 l  
171.780 139.665 l  
172.063 139.502 l  
172.346 139.351 l  
172.630 139.210 l  
172.913 139.077 l  
173.197 138.951 l  
173.480 138.831 l  
173.764 138.716 l  
174.047 138.607 l  
174.331 138.501 l  
174.614 138.399 l  
174.898 138.301 l  
175.181 138.206 l  
175.465 138.115 l  
175.748 138.026 l  
176.031 137.939 l  
176.315 137.855 l  
176.598 137.773 l  
176.882 137.693 l  
177.165 137.615 l  
177.449 137.539 l  
177.732 137.465 l  
178.016 137.393 l  
178.299 137.322 l  
178.583 137.253 l  
178.866 137.185 l  
179.150 137.118 l  
179.433 137.053 l  
179.717 136.989 l  
180.000 136.926 l  
180.283 136.865 l  
180.567 136.805 l  
180.850 136.745 l  
181.134 136.687 l  
181.417 136.630 l  
181.701 136.574 l  
181.984 136.518 l  
182.268 136.464 l  
182.551 136.411 l  
182.835 136.358 l  
183.118 136.306 l  
183.402 136.255 l  
183.685 136.205 l  
183.969 136.156 l  
184.252 136.107 l  
184.535 136.060 l  
184.819 136.012 l  
185.102 135.966 l  
185.386 135.920 l  
185.669 135.875 l  
185.953 135.831 l  
186.236 135.787 l  
186.520 135.744 l  
186.803 135.701 l  
187.087 135.659 l  
187.370 135.618 l  
187.654 135.577 l  
187.937 135.537 l  
188.220 135.497 l  
188.504 135.458 l  
188.787 135.419 l  
189.071 135.381 l  
189.354 135.344 l  
189.638 135.306 l  
189.921 135.270 l  
190.205 135.234 l  
190.488 135.198 l  
190.772 135.163 l  
191.055 135.128 l  
191.339 135.094 l  
191.622 135.060 l  
191.906 135.027 l  
192.189 134.994 l  
192.472 134.961 l  
192.756 134.929 l  
193.039 134.897 l  
193.323 134.866 l  
193.606 134.835 l  
193.890 134.805 l  
194.173 134.775 l  
194.457 134.745 l  
194.740 134.716 l  
195.024 134.687 l  
195.307 134.658 l  
195.591 134.630 l  
195.874 134.602 l  
196.157 134.575 l  
196.441 134.548 l  
196.724 134.521 l  
197.008 134.495 l  
197.291 134.469 l  
197.575 134.443 l  
197.858 134.417 l  
198.142 134.392 l  
198.425 134.368 l  
closepath 0.8 setgray fill 0 setgray 
1.5 setlinewidth 
0.000 141.732 m
28.346 141.732 l  
42.520 141.732 m
155.906 141.732 l  
170.079 141.732 m
198.425 141.732 l  
st
0.5 setlinewidth 
255.118 141.732 m
453.543 141.732 l  
st
425.197 141.732 m
425.188 142.845 l  
425.162 143.958 l  
425.118 145.071 l  
425.057 146.182 l  
424.978 147.292 l  
424.882 148.401 l  
424.769 149.509 l  
424.638 150.614 l  
424.490 151.717 l  
424.324 152.818 l  
424.142 153.916 l  
423.942 155.011 l  
423.724 156.103 l  
423.490 157.191 l  
423.239 158.276 l  
422.970 159.356 l  
422.685 160.432 l  
422.383 161.503 l  
422.064 162.570 l  
421.728 163.631 l  
421.376 164.687 l  
421.007 165.737 l  
420.622 166.782 l  
420.220 167.820 l  
419.802 168.852 l  
419.368 169.877 l  
418.918 170.895 l  
418.452 171.906 l  
417.970 172.909 l  
417.473 173.905 l  
416.960 174.893 l  
416.431 175.872 l  
415.887 176.844 l  
415.328 177.806 l  
414.754 178.760 l  
414.165 179.704 l  
413.561 180.639 l  
412.943 181.565 l  
412.310 182.481 l  
411.663 183.386 l  
411.001 184.282 l  
410.326 185.167 l  
409.637 186.041 l  
408.934 186.904 l  
408.218 187.756 l  
407.488 188.597 l  
406.746 189.426 l  
405.990 190.243 l  
405.222 191.049 l  
404.441 191.842 l  
403.647 192.623 l  
402.842 193.391 l  
402.025 194.147 l  
401.195 194.890 l  
400.355 195.619 l  
399.502 196.336 l  
398.639 197.038 l  
397.765 197.728 l  
396.880 198.403 l  
395.985 199.064 l  
395.079 199.711 l  
394.163 200.344 l  
393.238 200.963 l  
392.303 201.567 l  
391.358 202.156 l  
390.405 202.730 l  
389.442 203.289 l  
388.471 203.833 l  
387.491 204.361 l  
386.503 204.874 l  
385.508 205.372 l  
384.504 205.854 l  
383.493 206.320 l  
382.475 206.770 l  
381.450 207.204 l  
380.418 207.622 l  
379.380 208.024 l  
378.336 208.409 l  
377.285 208.778 l  
376.230 209.130 l  
375.168 209.466 l  
374.102 209.785 l  
373.030 210.087 l  
371.954 210.372 l  
370.874 210.640 l  
369.790 210.892 l  
368.701 211.126 l  
367.610 211.343 l  
366.515 211.543 l  
365.417 211.726 l  
364.316 211.891 l  
363.213 212.040 l  
362.107 212.170 l  
361.000 212.284 l  
359.891 212.380 l  
358.780 212.459 l  
357.669 212.520 l  
356.557 212.563 l  
355.444 212.590 l  
354.331 212.598 l  
353.218 212.590 l  
352.105 212.563 l  
350.992 212.520 l  
349.881 212.459 l  
348.771 212.380 l  
347.662 212.284 l  
346.554 212.170 l  
345.449 212.040 l  
344.346 211.891 l  
343.245 211.726 l  
342.147 211.543 l  
341.052 211.343 l  
339.960 211.126 l  
338.872 210.892 l  
337.787 210.640 l  
336.707 210.372 l  
335.631 210.087 l  
334.560 209.785 l  
333.493 209.466 l  
332.432 209.130 l  
331.376 208.778 l  
330.326 208.409 l  
329.281 208.024 l  
328.243 207.622 l  
327.211 207.204 l  
326.186 206.770 l  
325.168 206.320 l  
324.157 205.854 l  
323.154 205.372 l  
322.158 204.874 l  
321.170 204.361 l  
320.191 203.833 l  
319.219 203.289 l  
318.257 202.730 l  
317.303 202.156 l  
316.359 201.567 l  
315.424 200.963 l  
314.498 200.344 l  
313.582 199.711 l  
312.677 199.064 l  
311.781 198.403 l  
310.896 197.728 l  
310.022 197.038 l  
309.159 196.336 l  
308.307 195.619 l  
307.466 194.890 l  
306.637 194.147 l  
305.819 193.391 l  
305.014 192.623 l  
304.221 191.842 l  
303.440 191.049 l  
302.672 190.243 l  
301.916 189.426 l  
301.173 188.597 l  
300.444 187.756 l  
299.727 186.904 l  
299.025 186.041 l  
298.335 185.167 l  
297.660 184.282 l  
296.999 183.386 l  
296.352 182.481 l  
295.719 181.565 l  
295.100 180.639 l  
294.496 179.704 l  
293.907 178.760 l  
293.333 177.806 l  
292.774 176.844 l  
292.230 175.872 l  
291.702 174.893 l  
291.189 173.905 l  
290.691 172.909 l  
290.209 171.906 l  
289.743 170.895 l  
289.293 169.877 l  
288.859 168.852 l  
288.441 167.820 l  
288.039 166.782 l  
287.654 165.737 l  
287.285 164.687 l  
286.933 163.631 l  
286.597 162.570 l  
286.278 161.503 l  
285.976 160.432 l  
285.691 159.356 l  
285.423 158.276 l  
285.171 157.191 l  
284.937 156.103 l  
284.720 155.011 l  
284.520 153.916 l  
284.337 152.818 l  
284.172 151.717 l  
284.023 150.614 l  
283.893 149.509 l  
283.779 148.401 l  
283.683 147.292 l  
283.604 146.182 l  
283.543 145.071 l  
283.500 143.958 l  
283.473 142.845 l  
283.465 141.732 l  
283.473 140.619 l  
283.500 139.506 l  
283.543 138.394 l  
283.604 137.283 l  
283.683 136.172 l  
283.779 135.063 l  
283.893 133.956 l  
284.023 132.850 l  
284.172 131.747 l  
284.337 130.646 l  
284.520 129.548 l  
284.720 128.453 l  
284.937 127.362 l  
285.171 126.273 l  
285.423 125.189 l  
285.691 124.109 l  
285.976 123.033 l  
286.278 121.961 l  
286.597 120.895 l  
286.933 119.833 l  
287.285 118.778 l  
287.654 117.727 l  
288.039 116.683 l  
288.441 115.645 l  
288.859 114.613 l  
289.293 113.588 l  
289.743 112.570 l  
290.209 111.559 l  
290.691 110.555 l  
291.189 109.560 l  
291.702 108.572 l  
292.230 107.592 l  
292.774 106.621 l  
293.333 105.658 l  
293.907 104.705 l  
294.496 103.760 l  
295.100 102.825 l  
295.719 101.900 l  
296.352 100.984 l  
296.999 100.078 l  
297.660 99.183 l  
298.335 98.298 l  
299.025 97.424 l  
299.727 96.561 l  
300.444 95.708 l  
301.173 94.868 l  
301.916 94.038 l  
302.672 93.221 l  
303.440 92.416 l  
304.221 91.622 l  
305.014 90.841 l  
305.819 90.073 l  
306.637 89.317 l  
307.466 88.575 l  
308.307 87.845 l  
309.159 87.129 l  
310.022 86.426 l  
310.896 85.737 l  
311.781 85.062 l  
312.677 84.400 l  
313.582 83.753 l  
314.498 83.120 l  
315.424 82.502 l  
316.359 81.898 l  
317.303 81.309 l  
318.257 80.735 l  
319.219 80.176 l  
320.191 79.632 l  
321.170 79.103 l  
322.158 78.590 l  
323.154 78.093 l  
324.157 77.611 l  
325.168 77.145 l  
326.186 76.695 l  
327.211 76.261 l  
328.243 75.843 l  
329.281 75.441 l  
330.326 75.056 l  
331.376 74.687 l  
332.432 74.335 l  
333.493 73.999 l  
334.560 73.680 l  
335.631 73.378 l  
336.707 73.093 l  
337.787 72.824 l  
338.872 72.573 l  
339.960 72.339 l  
341.052 72.121 l  
342.147 71.921 l  
343.245 71.739 l  
344.346 71.573 l  
345.449 71.425 l  
346.554 71.294 l  
347.662 71.181 l  
348.771 71.085 l  
349.881 71.006 l  
350.992 70.945 l  
352.105 70.901 l  
353.218 70.875 l  
354.331 70.866 l  
355.444 70.875 l  
356.557 70.901 l  
357.669 70.945 l  
358.780 71.006 l  
359.891 71.085 l  
361.000 71.181 l  
362.107 71.294 l  
363.213 71.425 l  
364.316 71.573 l  
365.417 71.739 l  
366.515 71.921 l  
367.610 72.121 l  
368.701 72.339 l  
369.790 72.573 l  
370.874 72.824 l  
371.954 73.093 l  
373.030 73.378 l  
374.102 73.680 l  
375.168 73.999 l  
376.230 74.335 l  
377.285 74.687 l  
378.336 75.056 l  
379.380 75.441 l  
380.418 75.843 l  
381.450 76.261 l  
382.475 76.695 l  
383.493 77.145 l  
384.504 77.611 l  
385.508 78.093 l  
386.503 78.590 l  
387.491 79.103 l  
388.471 79.632 l  
389.442 80.176 l  
390.405 80.735 l  
391.358 81.309 l  
392.303 81.898 l  
393.238 82.502 l  
394.163 83.120 l  
395.079 83.753 l  
395.985 84.400 l  
396.880 85.062 l  
397.765 85.737 l  
398.639 86.426 l  
399.502 87.129 l  
400.355 87.845 l  
401.195 88.575 l  
402.025 89.317 l  
402.842 90.073 l  
403.647 90.841 l  
404.441 91.622 l  
405.222 92.416 l  
405.990 93.221 l  
406.746 94.038 l  
407.488 94.868 l  
408.218 95.708 l  
408.934 96.561 l  
409.637 97.424 l  
410.326 98.298 l  
411.001 99.183 l  
411.663 100.078 l  
412.310 100.984 l  
412.943 101.900 l  
413.561 102.825 l  
414.165 103.760 l  
414.754 104.705 l  
415.328 105.658 l  
415.887 106.621 l  
416.431 107.592 l  
416.960 108.572 l  
417.473 109.560 l  
417.970 110.555 l  
418.452 111.559 l  
418.918 112.570 l  
419.368 113.588 l  
419.802 114.613 l  
420.220 115.645 l  
420.622 116.683 l  
421.007 117.727 l  
421.376 118.778 l  
421.728 119.833 l  
422.064 120.895 l  
422.383 121.961 l  
422.685 123.033 l  
422.970 124.109 l  
423.239 125.189 l  
423.490 126.273 l  
423.724 127.362 l  
423.942 128.453 l  
424.142 129.548 l  
424.324 130.646 l  
424.490 131.747 l  
424.638 132.850 l  
424.769 133.956 l  
424.882 135.063 l  
424.978 136.172 l  
425.057 137.283 l  
425.118 138.394 l  
425.162 139.506 l  
425.188 140.619 l  
425.197 141.732 l  
404.441 191.842 m
5.669 0.000 rl  
-5.669 0.000 rl  
0.000 -5.669 rl  
0.000 5.669 rl  
st
1.5 setlinewidth 
255.118 141.732 m
283.465 141.732 l  
425.197 141.732 m
453.543 141.732 l  
st
st
showpage 

EndOfTheIncludedPostscriptMagicCookie
\closepsdump
\psdump{nmicf3.ps}%!
/st {stroke} def
/m {moveto} def
/rm {rmoveto} def
/l {lineto} def
/rl {rlineto} def
0.5 setlinewidth 
0.5 setlinewidth 
396.850 0.000 m
56.693 340.157 l  
113.386 396.850 l  
453.543 56.693 l  
closepath 0.9 setgray fill 0 setgray 
396.850 0.000 m
0.000 42.520 rl  
-14.173 14.173 rl  
-42.520 0.000 rl  
0.000 42.520 rl  
-14.173 14.173 rl  
-42.520 0.000 rl  
0.000 42.520 rl  
-14.173 14.173 rl  
-42.520 0.000 rl  
0.000 42.520 rl  
-14.173 14.173 rl  
-42.520 0.000 rl  
0.000 42.520 rl  
-14.173 14.173 rl  
-42.520 0.000 rl  
0.000 42.520 rl  
-14.173 14.173 rl  
-42.520 0.000 rl  
113.386 396.850 l  
0.000 -42.520 rl  
14.173 -14.173 rl  
42.520 0.000 rl  
0.000 -42.520 rl  
14.173 -14.173 rl  
42.520 0.000 rl  
0.000 -42.520 rl  
14.173 -14.173 rl  
42.520 0.000 rl  
0.000 -42.520 rl  
14.173 -14.173 rl  
42.520 0.000 rl  
0.000 -42.520 rl  
14.173 -14.173 rl  
42.520 0.000 rl  
0.000 -42.520 rl  
14.173 -14.173 rl  
42.520 0.000 rl  
closepath 0.8 setgray fill 0 setgray 
2 setlinewidth 
396.850 0.000 m
0.000 42.520 rl  
-14.173 14.173 rl  
-42.520 0.000 rl  
0.000 42.520 rl  
-14.173 14.173 rl  
-42.520 0.000 rl  
0.000 42.520 rl  
-14.173 14.173 rl  
-42.520 0.000 rl  
0.000 42.520 rl  
-14.173 14.173 rl  
-42.520 0.000 rl  
0.000 42.520 rl  
-14.173 14.173 rl  
-42.520 0.000 rl  
0.000 42.520 rl  
-14.173 14.173 rl  
-42.520 0.000 rl  
st
113.386 396.850 m
0.000 -42.520 rl  
14.173 -14.173 rl  
42.520 0.000 rl  
0.000 -42.520 rl  
14.173 -14.173 rl  
42.520 0.000 rl  
0.000 -42.520 rl  
14.173 -14.173 rl  
42.520 0.000 rl  
0.000 -42.520 rl  
14.173 -14.173 rl  
42.520 0.000 rl  
0.000 -42.520 rl  
14.173 -14.173 rl  
42.520 0.000 rl  
0.000 -42.520 rl  
14.173 -14.173 rl  
42.520 0.000 rl  
st
0.5 setlinewidth 
113.386 396.850 m
453.543 56.693 l  
st
56.693 340.157 m
396.850 0.000 l  
st
113.386 396.850 m
0.000 -113.386 rl  
113.386 0.000 rl  
0.000 -113.386 rl  
113.386 0.000 rl  
0.000 -113.386 rl  
113.386 0.000 rl  
st
56.693 340.157 m
113.386 0.000 rl  
0.000 -113.386 rl  
113.386 0.000 rl  
0.000 -113.386 rl  
113.386 0.000 rl  
0.000 -113.386 rl  
st
0.000 170.079 m
453.543 170.079 l  
-5.669 2.835 rl  
5.669 -2.835 rl  
-5.669 -2.835 rl  
5.669 2.835 rl  
226.772 0.000 m
226.772 396.850 l  
-2.835 -5.669 rl  
2.835 5.669 rl  
2.835 -5.669 rl  
-2.835 5.669 rl  
st
124.016 350.787 m
116.929 343.701 l  
0.000 2.835 rl  
0.000 -2.835 rl  
2.835 0.000 rl  
-2.835 0.000 rl  
st
180.709 294.094 m
173.622 287.008 l  
0.000 2.835 rl  
0.000 -2.835 rl  
2.835 0.000 rl  
-2.835 0.000 rl  
st
237.402 237.402 m
230.315 230.315 l  
0.000 2.835 rl  
0.000 -2.835 rl  
2.835 0.000 rl  
-2.835 0.000 rl  
st
294.094 180.709 m
287.008 173.622 l  
0.000 2.835 rl  
0.000 -2.835 rl  
2.835 0.000 rl  
-2.835 0.000 rl  
st
350.787 124.016 m
343.701 116.929 l  
0.000 2.835 rl  
0.000 -2.835 rl  
2.835 0.000 rl  
-2.835 0.000 rl  
st
407.480 67.323 m
400.394 60.236 l  
0.000 2.835 rl  
0.000 -2.835 rl  
2.835 0.000 rl  
-2.835 0.000 rl  
st
st
showpage 

EndOfTheIncludedPostscriptMagicCookie
\closepsdump
\psdump{nmicf4.ps}%!
/st {stroke} def
/m {moveto} def
/rm {rmoveto} def
/l {lineto} def
/rl {rlineto} def
0.5 setlinewidth 
141.732 0.000 m
141.732 283.465 l  
311.811 283.465 l  
311.811 0.000 l  
141.732 0.000 l  
closepath 0.8 setgray fill 0 setgray 
0.5 setlinewidth 
85.039 141.732 m
368.504 141.732 l  
-8.504 2.835 rl  
8.504 -2.835 rl  
-8.504 -2.835 rl  
8.504 2.835 rl  
st
226.772 0.000 m
226.772 283.465 l  
-2.835 -8.504 rl  
2.835 8.504 rl  
2.835 -8.504 rl  
-2.835 8.504 rl  
st
3 setlinewidth 
141.732 0.000 m
141.732 283.465 l  
311.811 0.000 m
311.811 283.465 l  
226.772 184.252 m
226.772 282.047 l  
st
st
showpage 

EndOfTheIncludedPostscriptMagicCookie
\closepsdump

\magnification\magstep1

\def\varcm{ truecm }

\centerline{\bf Microcausality and Energy-Positivity in all frames} 
\centerline{\bf imply Lorentz Invariance of dispersion laws}

\vskip 0.2cm 
\centerline { by }

\vskip 0.2cm 
\centerline { {\bf Jacques Bros$^1$} and {\bf Henri Epstein$^2$}} 

\vskip 0.2cm 
\centerline {${ }^1$ \bf Service de Physique Th\'eorique, CEA-Saclay,  
F-91191 Gif-sur-Yvette, France}

\centerline {${ }^2$ \bf Institut des Hautes Etudes Scientifiques,  
F-91440 Bures-sur-Yvette, France}

\vskip 0.5cm 
\noindent  
{\bf Abstract.}\  
 A new presentation of the Borchers-Buchholz result of the Lorentz-invariance 
of the 
energy-momentum spectrum in theories with broken Lorentz symmetry is given 
in terms of properties of the Green's functions of microcausal Bose and
Fermi-fields. 
Strong constraints based on complex geometry phenomenons are shown to 
result from the interplay of the basic 
principles of causality and stability in Quantum Field Theory:  
if microcausality and energy-positivity in all Lorentz frames 
are satisfied, then 
it is unavoidable that all stable particles of the theory   
be governed by Lorentz-invariant dispersion laws; in all the field sectors, 
discrete parts outside the continuum as well as the thresholds 
of the continuous parts of the energy-momentum spectrum, with possible holes
inside it,  are necessarily 
represented by mass-shell hyperboloids (or the light-cone). No violation of this
geometrical fact can be produced by   
spontaneous breaking of the Lorentz symmetry,  even if 
the field-theoretical framework is enlarged  
so as to include short-distance singularities
wilder than distributions.

\vskip 0.5cm

\centerline{\bf 1- Introduction}

\vskip 0.5cm

In a recent work [1], it has been advocated that 
the occurrence of spontaneous Lorentz and CPT violations in Quantum Field 
Theories governed by suitable {\sl non-local} Lagrangians can very well 
generate {\sl non-Lorentz-invariant dispersion laws} 
\footnote{${ }^{(1)}$}{We prefer keeping here
the terminology of 
``dispersion law'' (used traditionally e.g. in Thermal Quantum Field Theory) 
rather than adopting the new usage of ``dispersion relation'', which is of
course confusing 
in a domain where (Cauchy-type) dispersion relations relating the absorptive and
dispersive parts of Feynman-type amplitudes 
remain a basic tool of frequent use.}
{\sl which avoid the problems with stability and causality}.  
Such Lorentz violation effects produced at Planck scale might then in principle
be observed at 
lower energies in particle physics. In support of their claim, the authors of
[1]  
have produced examples of possible ``non-local models'' in which the quadratic
part of the Lagrangian corresponds to  
a dispersion law 
$p_0 = \omega(\vec p)$ enjoying the following properties:

a) The hypersurface ${\cal M}$ with equation  
$p_0 = \omega (\vec p)$ 
differs from a Lorentz-invariant mass shell hyperboloid, 

b) ${\cal M}$ is contained in the positive energy-momentum cone $\overline V^+ \
(p_0 \geq |\vec p|),$ 

c) For every momentum $\vec p$, the ``group velocity condition'' $|\partial
\omega (\vec p)| \leq 1$ holds, 
which means that ${\cal M}$ admits a space-like (or light-like) tangent
hyperplane at each of its points.  

\vskip 0.3cm

While condition b) expresses energy-positivity in all Lorentz frames, condition c)
ensures that all 
wave-packets satisfying the dispersion law $p_0 = \omega (\vec p)$ propagate
``essentially'' 
with a {\sl subluminal (or luminal) velocity};  
essentially means ``up to the quantum spreading of wave-packets, of the order of
the Planck constant'',  
as it is the case for the solutions of the Klein-Gordon and Dirac equations.  
However, in spite of this causal interpretation in terms of particle 
wave-packets, condition c) should by
no means 
be taken as a criterion of {\sl microcausality } for the underlying Quantum
Field Theory. 
Microcausality states that the commutator (resp. anticommutator)
$[\Phi(x),\Phi(x')]_{\mp}$ of 
a boson (resp.fermion) field $\Phi(x)$ should vanish in the whole region of 
{\sl relativistic spacelike separation} $\{(x,x');\  (x-x')^2 <0 \}.$ As we
shall see below,
the ``group velocity condition'' c) only appears as a necessary consequence of
microcausality,
but the converse is not true. 
This is why,
without underestimating the interest of the mentioned examples of [1] satisfying  
conditions a), b) and c), it appeared worthwhile to us to warn the 
field-theorist community that such examples cannot correspond to 
field models satisfying microcausality. In fact, at a high level  
of generality including field theories with local singularities 
of arbitrary strength (i.e. wilder than distributions) in spacetime, 
the requirement of microcausality represents such a strong constraint that,
when combined with energy positivity in all frames, it definitely implies the
following properties:  
 
 i) any dispersion law describing particles generated by the field 
is Lorentz invariant, namely the corresponding hypersurface ${\cal M}$ is a
sheet of 
hyperboloid with equation of the form $p_0 = \sqrt {{\vec p}^2 + m^2}$ (or the
light cone  
$\partial V^+$ if $m=0$).  

 ii) In all the sectors (or collision channels) of the 
space of states of the (interacting) field theory considered,   
the hypersurfaces which border the continuous part of the 
energy-momentum spectrum, including possible holes 
in the latter,
are also Lorentz-invariant, namely sheets of hyperboloid 
of the form $p_0 =\sqrt{{\vec p}^2 + M_i^2} $ (or the light-cone).  

\vskip 0.1cm
It is the purpose of the present paper to give a hopefully elementary
and self-contained proof of the
latter facts, which have been established long ago in a general, although
slightly different, 
framework by Borchers and Buchholz. 
As a matter of fact, the interest for the possible occurrence of
Lorentz-symmetry 
breaking is not new and it has already been a subject of deep investigation in
the 
framework of the basic principles of Quantum Field Theory (QFT): the latter two
properties
of Lorentz-invariance of the energy-momentum spectrum have indeed been proven   
in a paper by H.J. Borchers and D. Buchholz entitled 
``The Energy-Momentum Spectrum in Local Field Theories with Broken 
Lorentz-Symmetry''[2] completed by a paper by H.J. Borchers 
entitled ``Locality and covariance of the spectrum''[3] in the general framework
of 
Algebraic QFT (or ``Local Quantum Physics'') [4].  
In this deep analysis, generalizing similar results already obtained in [5] 
(see also [6] for a complete survey of the question), it was proven that 
the interplay of a weak form of microcausality, namely the commutativity of 
local observables attached to pairs of mutually space-like regions, together
with 
energy-positivity in all Lorentz frames was sufficient to produce a 
Lorentz-invariant shape of the energy-momentum spectrum, even if the
Lorentz-symmetry
was broken in the considered physical representation of the field observables. 
In view of the always vivid interest of the community for the possible
occurrence of 
some form of Lorentz-symmetry violation 
emerging from the spontaneous breaking at Planck scale 
of  a ``fundamental field or string theory''  
(see [1] and references therein), 
but also of its apparent unawareness of the results of [2,3],  
we think it useful to give a revival to these results in a way which 
we hope to be accessible to the current field-theorist reader.   
In fact, we wish 
to give here a new presentation of these unexpected properties of geometrical
nature
in energy-momentum space in terms of  
propagators and Green's functions of microcausal Bose and Fermi-fields. 
We shall thus avoid  using the less familiar formulation of
Algebraic QFT,  
and will focus on the contrary on the phenomena of complex geometry 
which play
a basic role in this
matter and still hold   
in a context of indefinite-metric (see our 
concluding remarks).   
  
Some preliminary comments on the exact content of properties i) and ii) 
(announced above) and on the relevant concepts and field-theoretical framework 
used in our approach are necessary. 

As in [2,3], the proof of properties i) and ii) which we shall give is 
of a general nature, i.e. non-perturbative 
and in fact independent of any Lagrangian formulation of the field model. 
However, our method is close to the preoccupations of [1] by its   
use of such basic objects as the transforms of causal (retarded and advanced)   
propagators of the fields in energy-momentum space. 
In the approach of [1], the dispersion laws of particles are always  
associated with given quadratic parts of field Lagrangians incorporating
explicit Lorentz-symmetry 
breaking coefficients of appropriate type. Such dispersion laws  
therefore correspond to particles which are ``elementary'' with respect to the
field 
introduced in the Lagrangian, namely they appear as associated with poles  
of the propagator of this field in energy-momentum space.  
To be complete, another case of   
dispersion laws also deserves to be considered, namely those which  
correspond to ``composite'' particles  
of the field: the latter appear as associated with poles of the four-point (or
higher n-point) 
functions of the field in energy-momentum space; for example, this is the case
for the 
hadronic particles if the fundamental fields are those of the standard model.

Here we shall show in detail the previously announced geometrical properties 
of Lorentz invariance for the  
poles of propagators (corresponding to the case considered in [1]) 
and we shall also indicate the 
derivation of the corresponding equally valid results for the poles of 
four-point (or n-point) functions. 
We shall thus be concerned with stable (elementary or composite) particles, 
corresponding to 
discrete parts of the spectrum, not embedded in the continuum. The case of
unstable particles corresponding to possible complex poles of the Green's
functions in 
unphysical sheets is excluded from our study.

We wish to stress that the somewhat surprising phenomenon of geometrical
Lorentz-invariance
produced in the present problem has to do with peculiar properties of  
complex geometry in several complex variables. This phenomenon occurs 
as soon as the retarded and advanced propagators 
(or more general $n-$point functions) have Fourier-Laplace 
transforms, holomorphic  in specific ``tube domains'' 
of complex energy-momentum space
which correspond in a very general way to the causal support properties of 
these functions in spacetime, expressing the microcausality of the fields. 
\footnote{${ }^{(2)}$}
{Such properties, which are also
closely related 
to the Jost-Lehmann-Dyson (JLD) formula for causal commutators[7], 
have been thoroughly exploited in
[2,3] precisely 
in the spirit of [7].} 
The traditional framework of QFT in which these {\sl analyticity properties
of Green's functions in complex energy-momentum space} have been derived 
(see [8,9,17] and references therein) is the Wightman-LSZ [18,19] axiomatic 
framework in which the various $n-$point functions of the fields are 
supposed to be tempered distributions. As an application of the   
theory of Fourier-Laplace transforms of tempered distributions 
with causal supports, making use of the notion of  
``boundary values 
of analytic functions in the sense of 
distributions'' (see e.g. [12]), the previously mentioned   
analyticity domains of the Green's functions in momentum space  
have been obtained {\sl together with polynomial increase properties 
of the functions in their respective domains}.  From the viewpoint of 
the Lagrangian formalism of QFT, this framework is supposed to fit with 
{\sl conventional (renormalizable) field models} whose Lagrangian 
only involves a finite number of derivatives of the fields.  
In the various examples of conventional type of [1] in which Lorentz-
symmetry-breaking terms have been introduced so as to produce a 
non-Lorentz-invariant dispersion law (condition a)), it was found that
the conditions b),c) listed at the beginning of this introduction
could not be satisfied simultaneously:  
this fact is indeed consistent 
with our property i) since (as shown below) 
the violation of the group-velocity condition c)   
implies the violation of microcausality.

We now emphasize the existence of enlarged frameworks of QFT, 
making use of the general mathematical setting of hyperfunction 
theory (see e.g. [20] and references therein, the first generalization 
in this direction being the theory of Jaffe fields [21]). 
In the hyperfunction setting, the retarded and advanced Green's 
functions in spacetime may have arbitrarily wild singularities;  
as we shall see below, this does not prevent their 
Fourier-Laplace transforms from being well-defined in the relevant 
complex domains of momentum space, although they no longer have 
a polynomial rate of 
increase at infinity in these domains.
The analyticity properties of these transforms still express (at a high degree
of generality) the microcausality property of such possible theories,
which is now defined 
in terms of appropriately chosen classes of test-functions.
It will then be clear from our geometrical proof of properties i) and ii)  
that this proof works for the class of {\sl all holomorphic functions} in the 
domains considered, and therefore  
that the loss of polynomial increase properties in such general 
field theories does not invalidate the results. 
From the viewpoint of 
the Lagrangian formalism of QFT, 
this very general framework 
of QFT (implying no limitation on the strength of short-distance singularities)
should cover  
(provided they exist!) all {\sl nonconventional models} whose Lagrangian 
involves an infinite number of derivatives of the fields.

\vskip 0.1cm
In our section 2, we shall first specify the basic analyticity properties of retarded
and advanced 
two-point functions in tube domains, which express microcausality in complexified
energy-momentum space; for completeness, we shall explain the result 
in elementary terms for the non-trivial case of 
QFT in the hyperfunction setting, in comparison with the standard result
for Wightman fields in the setting of tempered distributions.  
This sketch of the general case is presented here (in Sec. 2.1) for 
convincing the reader that the validity of 
analyticity in complex momentum space is 
not submitted to restrictions due to 
lack of convergence of the Fourier integrals;  
however, this matter is outside our main purpose which is 
the derivation of properties i) and ii) and it may be skipped 
without inconvenience by the reader.   
Then we describe the procedure through which information 
on the energy-momentum spectrum 
is encoded in this general approach in  complex momentum space. 
We then formulate 
three basic results of complex geometry, called Properties A, B and C, whose
physical 
consequences in terms of {\sl admissible dispersion laws} 
are derived in a straightforward way:
Property A explains why the velocity group condition c) of dispersion laws is
implied 
by microcausality under a weak requirement of energy-positivity.  
Properties B and C provide a proof of the previous statements (i) and ii)) 
for the dispersion laws of elementary particles and for the thresholds
(and possible holes) of the continuous spectrum,  
under the joint requirement of microcausality and energy-positivity in all
frames.
A complete proof of Properties A, B and C is given in this section. 
In section 3, it is shown that similar consequences of  
microcausality and (weak or strong) energy-positivity 
requirements can be formulated in terms of    
momentum-space analyticity properties  
of four-point (resp.  more generally $2n-$point)  
Green's functions established in [8,9] (resp. [17d),e)]).   
The exact counterparts of Properties A,B,C, called respectively 
A',B',C', are then described and these phenomena of complex geometry are shown
to imply
the corresponding statements (i) and ii)) for the dispersion laws 
of composite particles and
for the 
thresholds (and possible holes) of the continuous spectrum in the channels  
considered. 
Section 4 gives concluding remarks.

\vskip 0.5cm
\centerline{\bf 2 Shape of the energy-momentum spectral supports for the
two-point functions} 
\vskip 0.5cm

\centerline{\bf 2.1 Expressing microcausality in complex momentum space} 
\vskip 0.5cm

For simplicity, we first recall the {\sl basic analyticity property in complex 
momentum space} by using conventional Fourier integrals (as if the fields were
``good functions '' of $x$)
before explaining why this property still holds with a high degree of  
generality in the rigorous settings of tempered distributions and hyperfunctions,
corresponding respectively to conventional and nonconventional field theories.

Let $F^+(p)$ and $F^-(p)$ (with $p=(p_0,\vec p)$) be respectively the Fourier
transforms of 
the vacuum expectation values of the retarded and advanced (anti-)commutators of
a general
(fermion or boson) quantum field $\Phi(x)$:  
$$ F^+(p) = i\int {\rm e}^{ip\cdot x} \ \theta(x_0)\ <[\Phi({x\over 2}),
\Phi(-{x\over 2})]_{\pm}>\ dx_0
d\vec x,\ \ \ \ \eqno(1)$$ 
$$ F^-(p) = -i\int {\rm e}^{ip\cdot x} \ \theta(-x_0)\ <[\Phi({x\over 2}),
\Phi(-{x\over 2})]_{\pm}>\ dx_0
d\vec x.\ \ \ \ \eqno(1')$$ 
For writing the latter, we have assumed as usual that the space of states in
which the field is acting 
carries a representation of the group of spacetime translations and that the
field is 
invariant under this representation; energy and momentum operators are the
corresponding generators of
this group.  
It is of current use to exploit the analyticity properties of 
$F^+(p)$ and $F^-(p)$ respectively in the upper and lower half-planes of the
complexified energy
variable $p_0$. However, the postulate of microcausality for the field $\Phi(x)$
implies much more. 
In fact, it requires that the retarded and advanced propagators occurring under
the integrals at the 
r.h.s. of Eqs (1) and (1') have respectively their supports contained in the
closed forward   
and backward cones $\overline V^+ \ (x_0 \geq |\vec x|)$ and $\overline V^-\
(x_0 \leq -|\vec x|)$.  
It then follows that the integrals (1) and (1') remain convergent and define
analytic functions 
of the complex energy-momentum vector $k=p+iq$, still denoted by $F^+(k)$ and
$F^-(k)$,  
in the respective domains $T^+ \ ( p\  {\rm arbitrary} , \ q \in V^+)$ and  
$T^-\ ( p\  {\rm arbitrary},\ q \in V^-)$;\  $V^+ = -V^-$ is the open forward
cone: $q_0 > |\vec q|$.      
\ $T^+$ and $T^-$ are called the ``forward and backward tubes''; they  contain
respectively the upper and 
lower half-planes in all their one-dimensional sections by (complexified)
time-like straight 
lines, interpreted as energy variables in all possible Lorentz frames.
$F^{\pm}(k)$ are
the ``Fourier-Laplace transforms'' 
of the retarded and advanced propagators in complex energy-momentum space;  
their boundary values $F^{\pm}(p)$ on the reals from (respectively) 
$T^{\pm}$ are the Fourier transforms themselves of these propagators.  

In view of the local singularities of the fields, the previous derivation of 
this basic analyticity property has to be corrected under two respects
in realistic cases, namely
i) splitting in a meaningful way the ``(anti-)commutator
function'' 
$ C(x) = <[\Phi({x\over 2}),
\Phi(-{x\over 2})]_{\pm}> $ with support   
$\overline V^+  \cup \overline V^-$ into its retarded and advanced components 
$R(x)$ and $A(x)$
with respective supports 
$\overline V^+$ and $ \overline V^-$, 
in such a way that $C(x)= -i (R(x) - A(x)$)  
(this was done formally in Eqs (1) and (1') 
by using the Heaviside multipliers $\theta(\pm x_0)$),\   
and ii) giving a sense to the formal integrals of (1) and (1') when $p$ is replaced 
by $k= p+iq$ respectively taken in $T^+$ and $T^-$.  

\vskip 0.2cm
In the framework of Wightman-LSZ fields  where $C(x)$ is a tempered distribution, 
there exists a standard splitting procedure (see e.g. [9]) which defines 
$R(x)$ and $A(x)$ as tempered distributions with respective supports
$\overline V^+$ and $ \overline V^-$ (up to an ambiguity which is simply 
a linear combination of $\delta(x)$ with derivatives of the latter). 
Then in view of this support property the Fourier-Laplace exponential  
$ {\rm e}^{i(p+iq)\cdot x}$ is seen to be a good test-function, 
respectively for $R(x)$ provided $q$ is taken in 
$V^+$ and for $A(x)$ provided $q$ is taken in  
$ V^-$, which justifies the corresponding analyticity property of   
$F^{\pm}(p+iq)$ in $T^{\pm}$. Moreover these functions are proved to
have polynomial increase for $k=p+iq$ tending to infinity in 
$T^{\pm}$ and to be bounded by an inverse power of $|q|$ near the reals:
for $p$ real, $F^{\pm}(p)$ are then defined as distribution boundary values 
of $F^{\pm}(k)$ from the respective tube domains $T^{\pm}$. 
The Fourier transform $\tilde C(p)$ of $C(x)$ in the (L. Schwartz) sense 
of tempered distributions, which  
has to be a measure (for spectral reasons, see below), is then given by the following  
relation between distributions 
$\tilde C(p)= -i (F^+(p) - F^-(p))$.    

\vskip 0.2cm
In the enlarged framework of QFT where the definition of 
field operators is given in the sense of hyperfunction theory (see e.g. [20]), 
$C(x)$ is a hyperfunction on  $x-$space,  
namely a functional on an appropriate space $\cal A$ of analytic functions $\varphi(x)$. 
However, if one needs this concept in order to include arbitrarily wild local 
singularities in $x-$space, one may keep as a  basic requirement the fact that 
in momentum space, the ``spectral function'' 
$\tilde C(p)=     
<\tilde \Phi(p)\tilde \Phi(-p)> -  
<\tilde \Phi(-p),\tilde \Phi(p)>$ 
\footnote{${ }^{(3)}$}{Here the ``bracket notation'' in terms of operator products 
is used only for  
its suggestive content; no (infinite!) energy-momentum conservation
$\delta-$function 
is involved in it.}  
remains a distribution and even a measure in $p_0$, 
depending continuously on $\vec p$.  
This is still a consequence of the existence of a unitary   
representation of the spacetime translations 
acting on the field operators in the Hilbert space of states;   
more intuitively speaking,  if one excludes phenomena of 
infrared divergences, the contributions    
$\delta(p_0 -\omega(\vec p))$ associated with dispersion laws of the theory    
are still considered as dominant singularities in momentum space.   
However, $\tilde C(p)$ is no longer a {\sl tempered} distribution
(i.e. it may have nonpolynomial increase at infinity, in correspondence with 
the wild local singular behaviour of $C(x)$).   
Mathematically, this means that a reasonable space $\cal  A$ of test-functions 
is the space of functions $\varphi(x)$ whose Fourier transforms $\tilde \varphi(p)$ 
belong to the Schwartz space $\cal D$ (functions infinitely differentiable 
with compact support): $\cal A$ is a space of entire functions 
$\varphi(x+iy)$ which decrease more rapidly than any inverse power of $|x|$ 
in the reals and are bounded by linear exponentials of $y$. 
Then the Fourier transformation relating $C$ and $\tilde C$ is defined 
in the usual way by the formula $(C, \overline\varphi) = 
(\tilde C, \overline{\tilde \varphi})$. 
Typically, a hyperfunction $C(x)$ satisfying the 
previous requirements admits a ``concrete 
representative'' (defined up to an arbitrary entire function of $x$) 
which is a function $\hat C(z_0,\vec x)$, holomorphic with respect to 
$z_0= x_0 +i y_0$ in the cut-plane excluding the reals.
$\hat C$ must be polynomially bounded at infinity (corresponding to the 
distribution character of $\tilde C(p)$), but  
its behaviour near the reals may be arbitrarily wild, 
corresponding to the idea that the hyperfunction $C(x)$ is (in a 
heuristic sense) the ``jump of $\hat C$ across the reals''. 
What remains indeed concretely defined is the following integral  
(independent of $\eta$, $\eta >0$, for all $\varphi$ in ${\cal A}$ ): 
$$(C,\varphi) = \int d\vec x \int_{-\infty}^{+\infty} dx_0 
\ \left[\hat C(x_0 +i \eta, \vec x)
\ \varphi(x_0 +i \eta, \vec x)
\ -\ \hat C(x_0 -i \eta, \vec x)
\ \varphi(x_0 -i \eta, \vec x) \right].\ \ \ \ \eqno(2)$$ 

\vskip 0.2cm
\noindent
{\sl Microcausality and splitting:}   
\ the formulation of a support condition like microcausality cannot be 
done as in the case of distributions by using localized test-functions 
since $\cal A$ does not contain such functions. But making use of 
the representative $\hat C$ of $C$, one can formulate the 
microcausality condition by postulating that the jump of 
$\hat C$ across the reals vanishes in the complement of 
$\overline V^+  \cup \overline V^-$, or in other words that 
$\hat C(x_0, \vec x)$ is holomorphic there (in $x_0$).  
Now it can be shown that a splitting of the form 
$\hat C(z_0, \vec x) = -i (\hat R(z_0,\vec x) - \hat A(z_0,\vec x))$
can always be done (with an ambiguity consisting in 
functions carrying a singularity only at   
$(x_0, \vec x) =0$) in such a way that $\hat R$ and $\hat A$ be 
holomorphic in the cut $z_0$-plane (for all $\vec x$), polynomially bounded 
at infinity 
and such that the jumps of $\hat R$ and $\hat A$ across the reals  
vanish respectively in the complements of  
$\overline V^+$ and   
$\overline V^-$.  
The feasibility of this splitting which is not surprising for $|\vec x|\neq 0$, 
(since the ``retarded and advanced cuts'' are separated in that case) requires 
more mathematical knowledge for $\vec x =0$; such a decomposition belongs    
to a category called ``Cousin's problem'' which is standard in complex analysis.    
Applying the procedure of formula (2) for defining the hyperfunctions $R(x)$ 
and $A(x)$ with respective representatives $\hat R$ and $\hat A$, one thus obtains
for $C$ a satisfactory splitting of the form 
$C(x)= -i (R(x) - A(x))$.   

\vskip 0.2cm
\noindent
{\sl Analyticity in the tubes $T^{\pm}$:}   
\ applying the definition (2) to $R$ and $A$, and taking into account the location 
of the corresponding cuts for $\hat R$ and $\hat A$ in the corresponding formulae, 
one can perform a contour distortion argument which yields the following 
alternative forms:
$$(R,\varphi) = \int d\vec x \int_{\gamma^+(\vec x)} dz_0 
\ \hat R(z_0 , \vec x)
\ \varphi(z_0 , \vec x), 
\ \ \ \ \eqno(2')$$ 
$$(A,\varphi) = \int d\vec x \int_{\gamma^-(\vec x)} dz_0 
\ \hat A(z_0 , \vec x)
\ \varphi(z_0 , \vec x),  
\ \ \ \ \eqno(2'')$$ 
where $\gamma^+(\vec x)$ and  
$\gamma^-(\vec x)$ are folded contours which surround respectively the 
half-lines $x_0 \geq |\vec x|$ and  
$x_0 \leq -|\vec x|$ in the $z_0-$plane.   
One now notices that the entire function 
$ {\rm e}^{i(p_0+iq_0) z_0 - i(\vec p+ i\vec q)\cdot \vec x}$ is 
an acceptable  test-function
$\varphi$ in (2') if $(q_0,\vec q)$ is in $V^+$,  
and in (2'') if $(q_0,\vec q)$ is in $V^-$.   
One thus justifies again in this very general case the fact that 
$F^+(k) = (R, {\rm e}^{i k\cdot x})$ 
and $F^-(k) = (A, {\rm e}^{i k\cdot x})$ are 
respectively holomorphic in the tubes $T^+$ and $T^-$. 
Moreover, by taking into account the distance $\eta$ of the contours
$\gamma^{\pm}(\vec x)$ to the corresponding cuts in the $z_0-$plane,   
one would show that these Fourier-Laplace transforms 
of $R$ and $A$ now satisfy bounds of the following form in their 
respective tube domains:
$$ |F^{\pm}(k)| \leq M_{\eta}\  {\rm e}^{\eta (|p_0| + |q_0|)},$$ 
where the positive number $\eta$ can be given an arbitrary value 
(the dependence on $\eta$ of the corresponding constant $M_{\eta}$
being unspecified). 

\vskip 0.5cm

\centerline{\bf 2.2 A problem of analytic completion} 
\vskip 0.5cm

So {\sl in the sector generated by ``one-field vector-states'' of the form $\int
\varphi(x)\Phi(x) dx >$},  
with $\varphi$ varying in the relevant space of test-functions 
(corresponding to the field-theoretical framework which one wishes to consider),   
microcausality is fully expressed by the analyticity of the pair of functions
$(F^+,F^-)$   
in the corresponding domains $T^+, T^-$. 
Now any usable information on the 
support of the energy-momentum spectrum 
of the theory in this sector  
amounts to specifying an open subset ${\cal R}$ of the (real) 
energy-momentum space
in which the distributions 
$<\tilde \Phi(p)\tilde \Phi(-p)>$
and $ <\tilde \Phi(-p),\tilde \Phi(p)>$ vanish together.  
In fact, such a support property implies the
{\sl coincidence relation} 
$F^+_{|{\cal R}} = F^-_{|{\cal R}}$, since  
the expression 
$$ F^+(p) - F^-(p) = i\tilde C(p) = 
i\ <[\tilde \Phi(p),\tilde \Phi(-p)]_{\pm}> \ \ \ \ \eqno(3)$$
vanishes in $\cal R.$
It then follows from a standard theorem of complex analysis 
called the ``edge-of-the-wedge theorem'' 
(still valid in the case of holomorphic functions with 
distribution-like boundary values, 
see [10] and references therein),
that $F^+(k)$ and $F^-(k)$ then admit a {\sl common analytic continuation}
$F(k)$ which is 
analytic in the union of $T^+$, $T^-$ and of a complex neighborhood of the real
set ${\cal R}$;
in other words, $F^+$ and $F^-$ ``communicate analytically'' through ${\cal R}$,
as functions of the
set of complex variables $k= (k_0, \vec k)$.

It is one of the basic phenomena of Analysis and Geometry in several complex
variables 
that arbitrary (connected) subsets of complex space ${\bf C}^n$ are not in
general ``natural'' for the 
class of holomorphic functions: this means that for such a general subset
$\Sigma$, all the  
functions holomorphic in $\Sigma$ admit an analytic continuation in a common
larger  
domain $\hat \Sigma$, called the holomorphy envelope of $\Sigma$. This
phenomenon,
which does not exist in the single-variable case, involves exclusively   
geometrical properties of the set $\Sigma$ and the extension from $\Sigma$ to
$\hat \Sigma$
can always be done in principle by an appropriate use of the Cauchy integral
formula; 
this {\sl analytic completion procedure} presents a strong analogy with the
procedure of taking the 
convex hull $\check S$ of a subset $S$ in the ordinary real space ${\bf R}^n$,
the 
notion of a ``natural 
holomorphy domain'' in ${\bf C}^n$  being a certain generalisation of the
notion of ``convex domain'' in ${\bf R}^n$ (see e.g. [11,12] and references
therein). As a matter of fact, 
the most standard and useful result 
in this connection is the so-called ``tube theorem'' (see e.g. [12]) which we
shall apply below: 
{\sl Any domain $D$ in ${\bf C}^n$ which is ``tube-shaped'', i.e. of the form 
${\bf R}^n + i B$ 
admits a holomorphy envelope which is the tube 
$\hat D = {\bf R}^n + i \check B$, where $\check B$ is the convex hull of $B$ in
${\bf R}^n$.}   

It turns out that sets of the form $\Sigma_{\cal R} = T^+ \cup T^- \cup {\cal
R}$  
are not natural and that,
for various choices of ${\cal R}$ of physical interest,  
the corresponding holomorphy envelope $\hat \Sigma$ or parts of it can be
computed
and unexpectedly strong results then follow.
Cases when ${\cal R}$ itself can be extended to a larger real region $\hat {\cal
R}$ 
(namely $\hat{\cal R} =\hat \Sigma \cap {\bf R}^n  \ \supset \ {\cal R}$) 
are specially interesting, since they correspond   
to enlarging the region on which the ``spectral function'' 
$ <[\tilde \Phi(p),\tilde \Phi(-p)]_{\pm}>$ is proven to vanish, and therefore
to refining our 
information on the support of the distribution 
$<\tilde \Phi(p)\tilde \Phi(-p)>$, called {\sl ``spectral support''}.  
Properties A and C given below are precisely of this
type. Property B is a basic example of a holomorphy envelope for a domain 
${\Sigma}_{\cal R}$ which exactly corresponds to 
the case when energy-positivity is satisfied in all frames.  

\vskip 0.2cm
{\bf 2.3 Microcausality implies dispersion laws with subluminal velocities}  

\vskip 0.4cm
If energy-positivity is required to hold {\sl only in privileged frames, 
such as the laboratory frame and a set of frames which have  
small velocities with respect to the latter} \footnote{${ }^{(4)}$}{This refers 
to the notion of ``concordant
frames'' introduced in [1]}, 
there exists a maximal region $\hat{\cal R}$ of the form $-\omega(\vec p) < p_0
< \omega(\vec p)$ 
(with $\omega(\vec p) \geq \gamma |\vec p|$ for some positive constant $\gamma$)
in which the (anti-)commutator function   
$<[\tilde \Phi(p),\tilde \Phi(-p)]_{\pm}>$ vanishes. We claim that, {\sl due to
microcausality, 
the hypersurface with equation 
$p_0 = \omega(\vec p)$ is not arbitrary: it has to be a space-like
hypersurface.} 
In fact, the geometry of the relativistic light-cone is deeply involved in the
implications of  
microcausality, as it results from the following

{\sl Property A (``Double-cone theorem''):  

\noindent
Let ${\cal R}_{a,b}$ be a neighborhood (in real $p-$space) of a given 
time-like segment $]a,b[$ with end-points $a$ and $b$ 
($b$ in the future of $a$). Then any function $F(k)$ holomorphic in  
$\Sigma_{{\cal R}_{a,b}}$ admits an analytic continuation in a (complex) domain
which contains  
the real region 
$\diamond_a^b$, where   
$ \diamond_a^b$ is the ``double-cone'' defined as the set of all points $p$ such
that 
$p$ is in the future of $a$ and in the past of $b$.} 

{\sl Interpretation of Property A:}\

Let ${\cal M}$ 
($p_0 = \omega (\vec p)$)  
be the hypersurface bordering the vanishing region $\hat{\cal R}$  
of the (anti-)commutator function 
of a certain field theory satisfying microcausality and energy-positivity in
privileged frames.  
Then for each point 
$b=(\omega (\vec p), \vec p)$ in ${\cal M}$,  
there exists some 
interval of the form $ \omega (\vec p) -\epsilon < p_0 < \omega (\vec p)$ and
some open neighborhood 
${\cal R}_{a,b}$ of the time-like segment $]a,b[$ defined by this interval 
(i.e. $a\equiv (\omega (\vec p) - \epsilon, \vec p)$)  
which lies in $\hat{\cal R}$.  
It then follows from Property A that the propagator $F(k) $ of this theory has
to be analytic
in the full double-cone $\diamond_a^b$, and therefore that the corresponding
(anti-)commutator function must vanish in this double-cone: therefore,
$\diamond_a^b$ belongs to
$\hat {\cal R}$, 
and this argument holds for every point $b$ of ${\cal M}$,   
which shows that ${\cal M}$ has to be a spacelike hypersurface.  

Similarly, assume that the vanishing region $\hat {\cal R}$ is accompanied by
another pair of  
maximal vanishing regions $\hat {\cal R}_1^{\pm} $
of the form $\omega (\vec p) < |p_0| < \omega_1(\vec p)$
of the  (anti-)commutator function.  
Then $p_0 = \omega (\vec p)$ appears as the dispersion law of a particle
corresponding to a pole 
$Z(\vec p) \over 
k_0 - \omega (\vec p)$ of the propagator $F(k)$. 
So the previous argument shows in this case that both the hypersurface ${\cal
M}$ describing the 
dispersion law of the particle and the hypersurface ${\cal M}_1\ (p_0
=\omega_1(\vec p))$ 
bordering the region 
$\hat {\cal R}_1^+$ have to be spacelike.  
The argument extends of course to the case of any (ordered) set of dispersion
laws corresponding to 
several particles. 
{\sl Therefore, for every particle appearing with an energy gap in 
the propagator of the field considered, microcausality alone  
implies that condition c) (subluminal or luminal velocities) 
is satisfied by such a particle.}  

\newdimen\fixhoffset
\fixhoffset=\hsize \relax
\advance\fixhoffset by -16.000\varcm
\divide \fixhoffset by 2
\hbox{\kern\fixhoffset\vbox to 12.00 \varcm{\offinterlineskip
\def\point#1 #2 #3 {\rlap{\kern #1 \varcm
\raise #2 \varcm \hbox{#3}}}
\def\spot{{\kern -0.2em\lower.55ex\hbox{$\bullet$}}}
\vfill\includegraphics{nmicf1.ps}
\smash{\hbox to \hsize{%
\point 14.10 6.00 {$p_1$}
\point 8.20 11.50 {$p_0$}
\point 3.50 5.60 {$-1$}
\point 8.10 5.60 {0}
\point 12.10 5.60 {$1$}
\point 8.10 1.50 {$-1$}
\point 8.10 10.10 {$1$}
\point 9.50 10.85 {$\delta_\varepsilon$}
\point 9.82 9.50 {$\check h_\lambda\ \hbox{for}\ |\lambda-1| < \eta$}
\point 11.05 4.00 {$\check h_1$}
\point 5.89 2.30 {$\check h_\lambda$}
\point 3.60 4.10 {$\diamond$}
\hfill}}}\hfill}

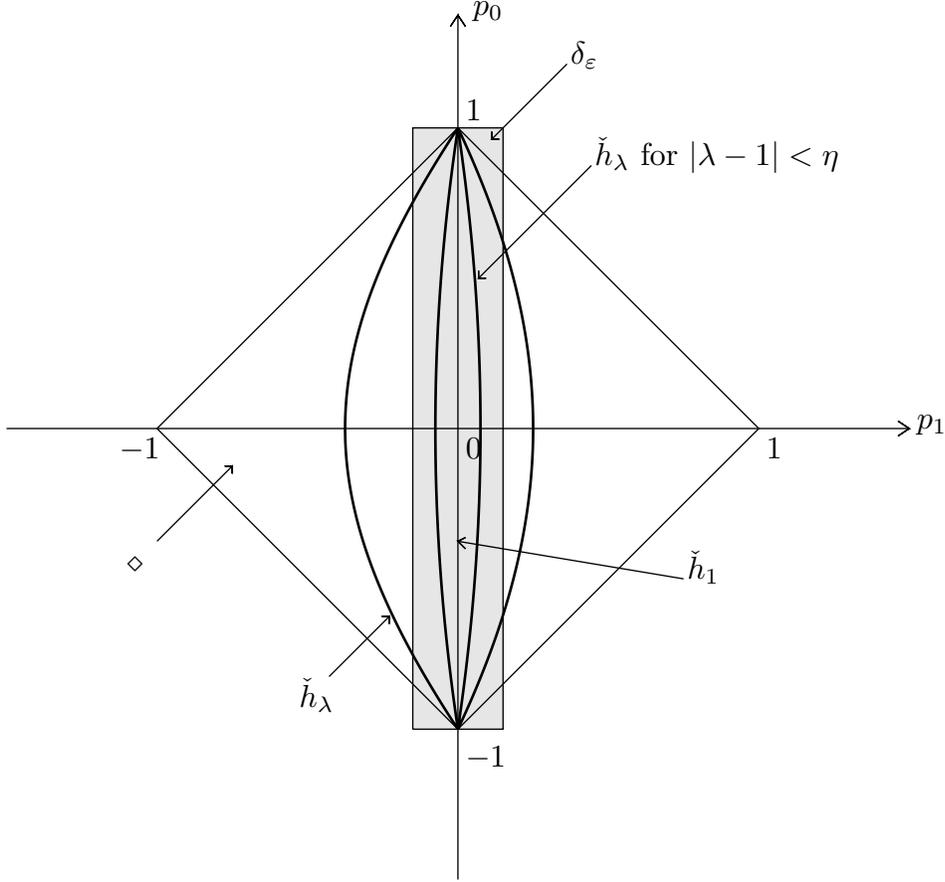
\centerline{Fig.~1. The ``double-cone'' $\diamond$ and the curves 
$\check h_\lambda$}

\vskip 0.2cm
{\sl Proof of Property A:}

This theorem, which can be seen as a generalisation of a similar property 
(corollary of the ``mean value Asgeirsson theorem'') for the solutions of 
the wave-equation [13], has been proved by Vladimirov [14] and by Borchers [15].
The main geometrical idea is displayed by treating a typical case in  
two-dimensional energy-momentum space with coordinates $(p_0,p_1)$.  
We take for $]a,b[$ the segment $\delta =]-1,+1[$ 
of the time axis and for ${\cal R}_{a,b}$ a thin rectangle $\delta_{\epsilon}$
of the form: 
$|p_0| < 1,\ |p_1| <\epsilon$. 
The tubes $T^+,T^-$ in the complexified space with coordinates  
$(k_0 = p_0 +i q_0,\ k_1= p_1 +i q_1)$ are defined respectively by the
conditions 
$q_0 +q_1>0,\ q_0-q_1 >0$ 
and 
$q_0 +q_1<0,\ q_0-q_1<0,$ 
and we shall show that the real region obtained by analytic 
completion of $T^+ \cup T^- \cup \delta_{\epsilon}$ contains the ``double-cone''
$\diamond$ (a square in this case!) defined by the inequalities: $|p_0 - p_1|
<1,\  |p_0 + p_1| <1.$   
One introduces the family of complex curves $h_{\lambda}$ with equation
$[k_0^2 -(k_1-1)^2]=  
\lambda [k_0^2 -(k_1+1)^2]$, 
where the parameter $\lambda$ varies in a 
complex neighborhood ${\cal V}$ of the real interval $]0,+\infty[$. 
Except for $h_1$ which is the (complexified) $p_0-$axis,  
all these curves are hyperbolae, and $\diamond $ is generated by the (real) arcs
$\breve h_{\lambda}$ of $h_{\lambda}$ parametrized by $-1 <p_0 < 1$ (with $|p_1|
<1$)  
when $\lambda $ varies from $0$ to $+\infty$; in a subinterval  
of the form $|\lambda -1| < \eta$ (for some $\eta$ 
determined by $\epsilon$),  
$\breve h_{\lambda}$ remains inside  
the rectangle $\delta_{\epsilon}$ (see fig 1). 

One then checks that for any function $F(k_0,k_1) $ holomorphic 
in $T^+ \cup T^- \cup {\delta_{\epsilon}}$  
the change of complex 
variables $(k_0,k_1) \to (k_0, \lambda)$ is admissible. It allows one to define 
$\underline F(k_0,\lambda) = F(k_0, k_1(k_0,\lambda)) $ as an analytic function 
for $\lambda $ varying in ${\cal V}$  and $k_0$ varying in 
a ring-shaped domain $D_{\lambda}$, which surrounds and {\sl excludes
a neighborhood of a real interval of the form }   
$-1 +\alpha (\lambda) \leq p_0 \leq 1- \alpha(\lambda)$ (fig 2a). 
This comes from the fact that for $0 < \lambda < +\infty$, the full upper (resp.
lower) 
half-plane in the variable $k_0$ represents a set of 
points $(k_0,k_1)$ of $h_{\lambda}$ in $T^+$ (resp. $T^-$) 
\footnote{${ }^{(5)}$}{To see this, one can e.g. rewrite the equation of
$h_{\lambda}$ as follows:
${U-1 \over U+1}=\lambda\  {V-1\over V+1}$ with $U= k_0 +k_1, \ V= k_0 -k_1,$ 
which entails (for 
$\lambda >0$)  
the condition $\Im m\, U \times \Im m\, V >0$, and therefore the fact that 
all complex points $(k_0,k_1) \equiv (U,V)$ in $h_{\lambda}$ belong either to 
$T^+$ or to $T^-$ according to whether $\Im m\, k_0 \equiv {1\over 2}(\Im m\, U
+ \Im m\, V)$ is 
positive or negative.}
and that these two half-planes are connected by small real intervals       
$]-1, -1+ \alpha[$, $]1- \alpha, 1[$ which represent points in
$\delta_{\epsilon}$.  
Moreover, for   
$1-\eta < \lambda <1+ \eta$ {\sl the full unit disk} $|k_0| <1$ is in the
analyticity domain 
of $\underline F$ (fig 2b), the corresponding arcs  
$\breve h_{\lambda}$ being all contained in   
$\delta_{\epsilon}$.

\vskip 0.10 truecm 
\newdimen\fixhoffset
\fixhoffset=\hsize \relax
\advance\fixhoffset by -16.000\varcm
\divide \fixhoffset by 2
\hbox{\kern\fixhoffset\vbox to 10.00 \varcm{\offinterlineskip
\def\point#1 #2 #3 {\rlap{\kern #1 \varcm
\raise #2 \varcm \hbox{#3}}}
\def\spot{{\kern -0.2em\lower.55ex\hbox{$\bullet$}}}
\vfill\includegraphics{nmicf2.ps}
\smash{\hbox to \hsize{%
\point 3.50 5.00 {\spot}
\point 0.50 4.50 {$-1$}
\point 6.10 4.50 {$1$}
\point 3.50 4.50 {0}
\point 1.30 4.50 {$\scriptstyle -1+\alpha(\lambda)$}
\point 4.70 4.50 {$\scriptstyle 1-\alpha(\lambda)$}
\point 1.50 1.00 {$a)\ \lambda\ \hbox{arbitrary in}\ {\cal V}$}
\point 12.50 5.00 {\spot}
\point 9.50 4.50 {$-1$}
\point 15.10 4.50 {$1$}
\point 12.50 4.50 {0}
\point 10.50 1.00 {$b)\ 1-\eta < \lambda\ < 1+\eta$}
\hfill}}}\hfill}

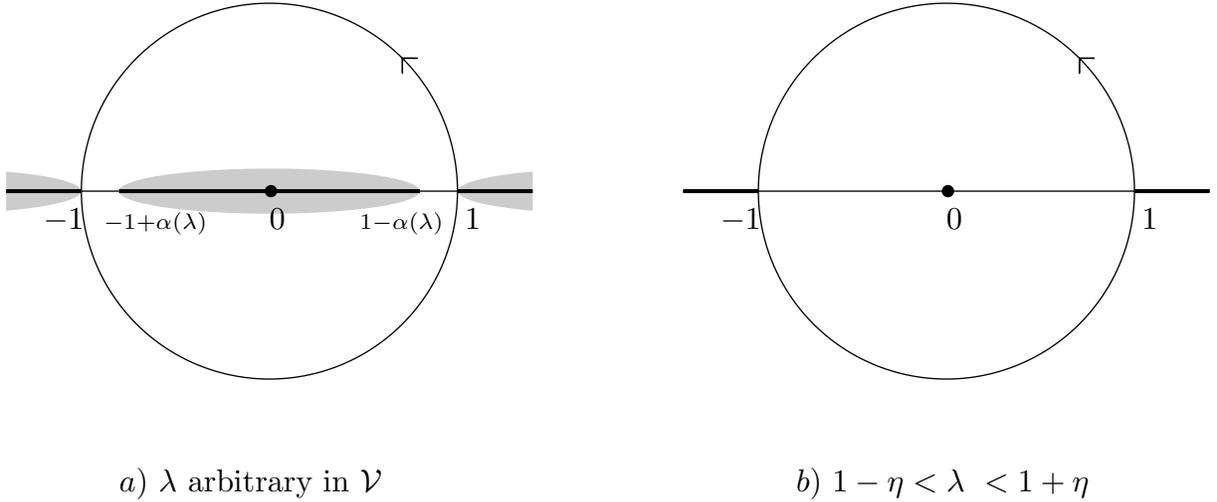
\centerline{Fig.~2. Initial analyticity domains of 
$\underline F(k_0,\ \lambda)$}
\centerline{in the $k_0$-plane}

\noindent
Now consider the Cauchy integral 
$$ I(k_0, \lambda) = {1\over 2i\pi} \oint_{\gamma_{\lambda}} 
{{\underline F}(k'_0,\lambda) \over k'_0 - k_0} dk'_0,$$ 
where $\gamma_{\lambda}$ has its support in $D_{\lambda}$, 
is homotopic to the (anticlockwise) unit circle and equal to it
for $\lambda = 1$.  
In view of the 
latter analyticity property of $\underline F$, one has  
$ I(k_0, \lambda) = \underline F(k_0, \lambda)$  
for $1-\eta < \lambda < 1+\eta$ and therefore  
$ I(k_0, \lambda) $ provides an analytic continuation of  
$ \underline F(k_0, \lambda)$ itself inside the full disk 
bordered by $\gamma_{\lambda}$ 
{and therefore on 
the real interval $]-1,+1[$ of the variable $k_0$,   
for all $\lambda $ in the interval $]o,+\infty[$.}   
By coming back to the original variables $(k_0,k_1)$, this shows that $F$ admits
an 
analytic continuation in the full region $\diamond$.
In the most general version of the theorem in two dimensions,  
the neighborhood ${\cal R}_{a,b}$ of the given time-like segment $]a,b[$ is
considered as a
union of rectangles of the previous $\delta_{\epsilon}-$type, whose thickness
$\epsilon$ 
tends to zero while they tend to $]a,b[$: the double-cone (or square) 
${\diamond_a^b}_{|d=2}$ is then clearly 
obtained as the union of the corresponding squares ${\diamond}$ obtained in the
previous 
procedure of analytic completion . Finally   
the proof of the theorem in the $d-$dimensional case is obtained by 
applying the two-dimensional result in all the planar sections passing
by $a$ and $b$, since i) the two-dimensional sections  of the tubes $T^{\pm}$
are 
the corresponding tubes of the (complexified) planar sections, and   
ii) $\diamond_a^b$ is generated by the union of all double-cones of 
the previous type ${\diamond_a^b}_{|d=2}$ in these planar sections.

\vskip 0.5cm
{\bf 2.4 Microcausality and energy-positivity in all frames imply Lorentz
invariant   
spectral supports} 
 
A basic implication of microcausality together with energy-positivity in all
frames is the 
fact that propagators $F(k)$ of the underlying fields have to be {\sl
holomorphic in a  
domain which is invariant under all complex Lorentz transformations}, even if
these 
propagators are not Lorentz invariant functions due to the fact that the Lorentz
symmetry is broken in the  
representation of the fields under consideration. The key property which is at
the origin 
of this peculiarity is the 
following

\vskip 0.2cm
{\sl Property B (``K\"allen-Lehmann domain''): 

\noindent
Let ${\cal R}= {\cal R}_0$ be the set of all 
space-like energy-momentum vectors $p= (p_0, \vec p):\   
|p_0| < |\vec p|$. Then  
any function $F(k)$ holomorphic in  
$\Sigma_{{\cal R}_0} =T^+ \cup T^-\cup {\cal R}_0$ admits an analytic
continuation in 
the domain $\hat \Sigma_{{\cal R}_0} $ which is the set of all 
{\sl complex} vectors $k= (k_0, \vec k)$ 
such that $k^2 \equiv k_0^2 - {\vec k}^2 $ is different from any positive number
and from zero.}  

\vskip 0.1cm
{\sl Interpretation of Property B:}\ \  

Energy-positivity in all Lorentz frames implies that the distribution 
$<\tilde \Phi(p)\tilde \Phi(-p)>$ vanishes in the complement of ${\overline
V}^+$   
and therefore, in view of (3),     
that the coincidence relation   
$F^+_{|{\cal R}_0} = F^-_{|{\cal R}_0}$ is satisfied .   
Property B then implies the analyticity of the propagator $F(k)$ in the full
``cut-domain'' 
$\hat \Sigma_{{\cal R}_0} $.
Our denomination of  
``K\"allen-Lehmann domain'' for the latter     
is motivated by the fact that 
in the usual case when Lorentz invariance (or covariance) of the field is
postulated, 
the analyticity domain 
$\hat \Sigma_{{\cal R}_0} $ is directly  
obtained as a byproduct of the 
K\"allen-Lehmann integral representation of the propagator 
$$ F(k) \equiv {\underline F}(k^2)  =
\int_0^{\infty} {\rho (\sigma)\over \sigma - k^2} d\sigma,$$
since the image of 
$\hat \Sigma_{{\cal R}_0} $
in the variable $k^2$ is   
the usual cut-plane domain ${\bf C} \setminus {\bf R}^+.$ 
Here, however, this 
Lorentz-invariant domain (considered in the full complex $k-$space) 
is obtained {\sl without any assumption
of Lorentz covariance  and of boundedness of the functions}, 
but purely on the basis of microcausality and energy-positivity.  

\vskip 0.1cm
Moreover, one will  show that any further information on the 
spectral support which is superimposed to the conditions of Property B 
implies the Lorentz-invariant
shape of all the components of the spectral support together with 
the invariance under complex Lorentz transformations
of the corresponding analyticity domain of the propagator. 
This is the purpose of the following property, whose statement 
in the present form is valid for any
spacetime dimension $d \geq 3$; we postpone to the proof the 
corresponding statement for the two-dimensional case, which requires  
a little more care in view of the decomposition of the light-cone 
into two straight-lines (the so-called ``left and right-movers'').  

\vskip 0.2cm
{\sl Property C (Lorentz-invariance of the borders of the spectral supports);
case $d\geq 3$: 

\noindent
If $\cal R$ is any real open set, not necessarily connected, containing ${\cal
R}_0$ then 
every function $F(k)$ holomorphic in 
$\Sigma_{\cal R} = T^+ \cup T^- \cup {\cal R}$ admits an  
analytic continuation in the (Lorentz-invariant) 
set $\hat {\cal R}$ of all real vectors $p$ 
whose Minkowskian norm $p^2$ has a value already taken at some vector in  ${\cal
R}$. 
Moreover the domain
$\hat \Sigma_{\cal R}$ in which every  
such function $F(k)$  can be analytically continued 
is the set of vectors $k$ such that $k^2$ takes all possible complex
values and all real values taken by $p^2$ when $p$ varies in ${\cal R}$.}  

\vskip 0.2cm
{\sl Interpretation of Property C:}\ \  

It is easy to see that Property $C$ (in its first part) implies that if
microcausality and
energy-positivity are satisfied, then the most general type of 
set $\hat {\cal R}$ where the (anti-)\break commutator function 
(3) has to vanish is a set composed of one distinguished region ${\cal R}_{M_0}$
of the  
form $-\infty < p^2 < {M_0}^2$, with ${M_0} \geq 0$ and of zero, one   
or several disjoint Lorentz-invariant regions of the form 
${M'}_i^2 < p^2 < {M}_{i}^2,$ where 
${M'}_1 \geq M_0 $ and 
${M'}_{i+1} \geq {M}_i,\ \ i= 1,\ldots, l-1,\ $
${M}_l \leq \infty$.
This implies in turn that the support of  
$<\tilde \Phi(p)\tilde \Phi(-p)>$ is exactly the union of all the ``thick (or
thin) 
hyperbolic shells'' defined by  
${M}_i^2 \leq p^2 \leq {M'}_{i+1}^2,\ p_0 \geq 0,$ ($i=0,1,\ldots,l-1$), $p^2
\geq M_l^2$   
(and of the origin if    
$<\tilde \Phi(p))> \not = 0$). 
The equality case  $M_i = {M'}_{i+1}$ corresponds to some ``thin shell'' $p^2=
M_i^2$. 
This thin shell situation occurs precisely    
when the distribution 
$<\tilde \Phi(p)\tilde \Phi(-p)>$ 
describes a particle with dispersion law $p_0 = \sqrt{ {\vec p}^2 + M_i^2 }.$   
No possibility is left for a Lorentz-symmetry breaking dispersion law. 
(Note that in this argument, 
the positivity of the Hilbert-space norm, 
implying the fact that the previous distribution is 
a positive measure factoring out 
a $\delta (p^2 - M_i^2)$, has not been used).

\vskip 0.2cm
The proofs of Properties B and C given below are based on purely geometrical
arguments.   
Both of them  rely on a standard analytic completion procedure of geometrical
type, namely  
the ``tube theorem'' (stated at the beginning of this section); apart from the  
recourse to this piece of knowledge in complex geometry, 
these proofs are completely self-contained.    
The analytic completion procedure is actually at work in the two-dimensional
case, which we treat 
at first, while the general $d-$ dimensional case will be reducible to the
latter.

\vskip 0.2cm
For the two-dimensional case, Property C must be properly restated as follows:

\vskip 0.2cm
{\sl Property C (Lorentz-invariance of the borders of the spectral supports);
case $d=2$: 

\noindent
If $\cal R$ is any real open set, not necessarily connected, containing ${\cal
R}_0$ then 
every function $F(k)$ holomorphic in 
$\Sigma_{\cal R} = T^+ \cup T^- \cup {\cal R}$ admits an  
analytic continuation in the (Lorentz-invariant) set 
$\hat \Sigma_{\cal R}$ obtained by adding to 
$\hat \Sigma_{{\cal R}_0}$  
the set of all (real or complex) vectors $k$ obtained by the action of 
real or complex Lorentz transformations  
on all vectors in ${\cal R}$. }   

\vskip 0.1cm
We note that in the $d-$dimensional case, the latter version of Property C is
equivalent to the 
former. In fact, the set of all vectors $k$ obtained 
from a given vector $p =\underline p \neq 0$ in ${\cal R}$ by real or 
complex Lorentz transformations is the full complexified hyperboloid $k^2 =
\underline p^2$ if 
$\underline p^2 \not= 0$ or the full complexified light cone $k^2=0$ if
$\underline p^2=0$. 
However in the two-dimensional case, the latter
statement differs from the former if ${\cal R}$ contains vectors $\underline p$ 
such that $\underline p^2=0$. 
In that case, the set of vectors $k$ obtained from such a vector $\underline p$
by the action of  
real or complex Lorentz transformations {\sl is not the full light cone} but
only the complexified line 
of left or right-movers 
which the given vector $\underline p$ itself belongs to. In other words, one of 
these two lines may very well be a singular set of the propagator, and therefore
contribute 
to the spectral support, although the other line doesn't; 
in such a case the parity symmetry of the 
spectral support is then  
broken but its Lorentz invariance is still preserved.   

{\sl Proof of Properties B and C in the two-dimensional case:}

We here consider the case when $k= (k_0,k_1)$ varies in ${\bf C}^2$,  
corresponding to two-dimensional field-theory. 
In the complex variables $(U= k_0 + k_1,\ V= k_0- k_1),$ the domains $T^{\pm}$
are described as 
$T^+:\ \Im m\, U >0, \ \Im m\, V >0,$ \  
$T^-:\ \Im m\, U <0, \ \Im m\, V <0,$  and ${\cal R}_0$ is the real set: $p^2 =
UV <0.$ 
Let us then pass to the logarithmic variables $u = \log U,\ v = \log V$ and use
the 
fact that any function $F(k)\equiv F(U,V) = F({\rm e}^u, {\rm e}^v)\equiv
f(u,v)$  
is holomorphic and 
$2\pi-$periodic with respect to the variables $u$ and $v$ in the image of 
$T^+ \cup T^- \cup {\cal R}_0$ in the space of these variables.   
One easily sees that 
the domain $T^+$ is one-to-one mapped (periodically) onto each one of the
following (tube-shaped) 
domains $\Theta^+_l = {\bf R}^2 + i B^+_l$ ($l$ integer) where $B^+_l$ is the
square $\ 0<\Im m\, u -2l\pi< \pi,\ 0 <\Im m\, v +2l\pi<\pi$ and similarly 
for $T^-$ onto each one of the  
domains $\Theta^-_l = {\bf R}^2 + i B^-_l$ ($l$ integer) where $B^-_l$ is the
square $\ -\pi<\Im m\, u -2l\pi< 0,\ \pi <\Im m\, v +2l\pi<2\pi$.   
As seen on fig 3, the set of all squares $B^+_l$ and $B^-_l$ form a connected
set {\sl if one adds to them
the common boundary vertices} represented by all the points 
$b_{l\over 2} =(\Im m\, u = l\pi,\ \Im m\, v= (-l+1)\pi) $, with $l$ integer.
But  
as one easily checks, the sets $\theta_{l\over 2} = {\bf R}^2 +i b_{l\over 2}$ 
belong precisely to the (periodic) image 
of the set ${\cal R}_0$ ($UV= {\rm e}^{u+v}<0;\ \ {\rm e}^u,\ {\rm e}^v $\
real).  
The function $f(u,v)$ is therefore holomorphic  
in the union of all the tube-shaped sets $\Theta^+_l,\ \Theta^-_l$ and
$\theta_{l\over 2}$ and 
even (in view of the invariance of this edge-of -the-wedge configuration  
by all real translations in ${\bf R}^2$) in a {\sl connected open tube} $\Theta
={\bf R}^2 +i B$ such that 
$B$ is the union of all 
sets $B^+_l,\ B^-_l$ together with {\sl open neighborhoods} of all the points
$b_{l\over 2}$.   
Then in view of the tube theorem, $f(u,v)$ admits a ($2\pi-$periodic) analytic
continuation in 
the tube $\hat \Theta = {\bf R}^2 +i \check B$, where  
$\check B$, namely the convex hull of $B$, is (as shown by fig 3) the 
domain $\check B: \ 0< \Im m\, u+ \Im m\, v < 2\pi$. 
$F(k)$ therefore admits an analytic continuation in the inverse image of    
the tube $\hat \Theta$  
in the original variables, which is the set of all $k\equiv (U,V)$ such that
$0< \arg U + \arg V \equiv \arg k^2 <2\pi, $  namely  
the domain $\hat \Sigma_{{\cal R}_0}$ described in Property B. 

\vskip 2 truecm 
\newdimen\fixhoffset
\fixhoffset=\hsize \relax
\advance\fixhoffset by -16.000\varcm
\divide \fixhoffset by 2
\hbox{\kern\fixhoffset\vbox to 14.00 \varcm{\offinterlineskip
\def\point#1 #2 #3 {\rlap{\kern #1 \varcm
\raise #2 \varcm \hbox{#3}}}
\def\spot{{\kern -0.2em\lower.55ex\hbox{$\bullet$}}}
\vfill\includegraphics{nmicf3.ps}
\smash{\hbox to \hsize{%
\point 4.42 12.43 {$b_{-1}$}
\point 6.42 10.43 {$b_{-1/2}$}
\point 8.43 8.43 {$b_{0}$}
\point 10.43 6.42 {$b_{1/2}$}
\point 12.43 4.42 {$b_{1}$}
\point 14.43 2.42 {$b_{3/2}$}
\point 2.50 12.50 {$B^{-}_{-1}$}
\point 4.50 10.50 {$B^{+}_{-1}$}
\point 6.50 8.50 {$B^{-}_{0}$}
\point 8.50 6.50 {$B^{+}_{0}$}
\point 10.50 4.50 {$B^{-}_{1}$}
\point 12.50 2.50 {$B^{+}_{1}$}
\point 14.50 0.50 {$B^{-}_{2}$}
\point 7.50 5.50 {0}
\point 15.00 5.50 {$\Im u$}
\point 8.20 13.50 {$\Im v$}
\hfill}}}\hfill}

\vskip 0.25 truecm
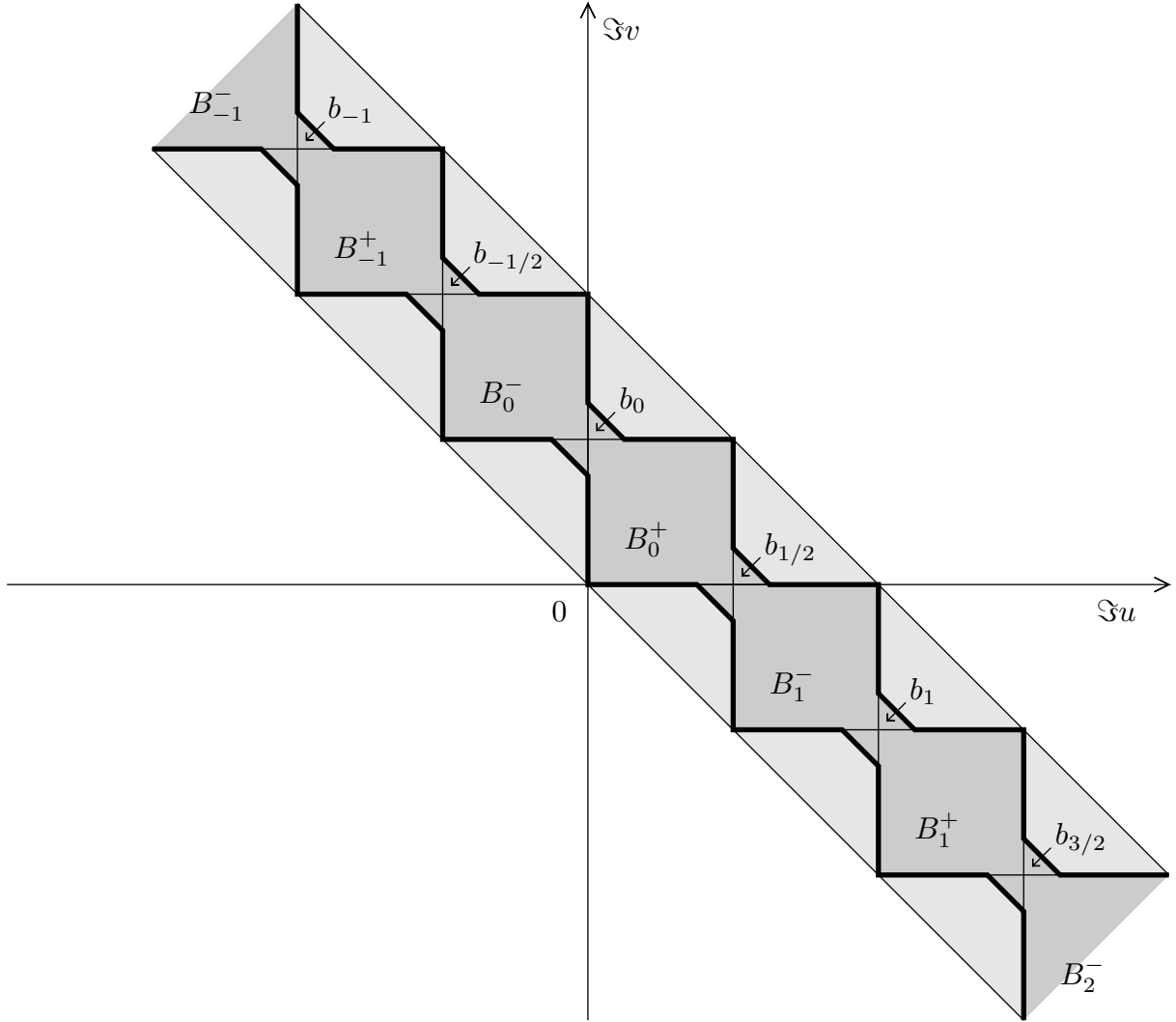
\centerline{Fig.~3. The set $B$ (dark gray)}
\centerline{and its convex hull $\check B$ (light gray)}

The domain $\hat \Sigma_{{\cal R}_0}$ can also be seen as the union of all
complex hyperbolae $h_{\zeta}$ 
in ${\bf C}^2$ with equation $k^2 = UV =\zeta$ such that $\zeta$ belongs to the
cut-plane   
${\bf C} \setminus [0,\infty[.$   
Let us now assume that in addition to   
${\cal R}_0$, the set ${\cal R}$ contains 
a given point $\underline p= (\underline U,\underline V)$ with 
$\underline p^2 = \underline \zeta \geq 0$. To be specific, consider the case
when one has: $\underline U >0$ and $\underline V \geq 0$ and put  
$\underline U= {\rm e}^{\underline t}\ >0,
\underline V= \underline \zeta {\rm e}^{-\underline t}\ \geq 0$, with $
\underline t$ real;   
the remaining cases would be treated similarly by i) exchanging 
the roles of $U$ and $V$ and ii)  changing $(U,V)$ into $(-U,-V)$ in the
following.  
We now use the fact that any 
function $F(k) \equiv F(U,V)$ analytic in $\Sigma_{\cal R} =T^+ \cup T^- \cup
{\cal R}$   
is analytic in a complex neighborhood of $\underline p$ 
and therefore in particular in a set of the form  
${\cal N}(\underline p) =\{ k=(U,V);\ U={\rm e}^t,\ V=\zeta {\rm e}^{-t};\
(\zeta,t) \in S_1 \}$, 
where $S_1 = \{(\zeta,t);  
\underline \zeta -\epsilon <\zeta <\underline \zeta + \epsilon,\   
|t-\underline t| < \rho \}. $   
It also follows from Property B that the image $G$ of such a function $F(U,V)$
in the space of
complex variables $(\zeta, t)$, namely $G(\zeta,t) \equiv F({\rm e}^t, \zeta
{\rm e}^{-t})$, is  
analytic in the set $S_2 =  \{ (\zeta,t);\ |\zeta -\underline \zeta| < \epsilon,
\  
\Im m\, \zeta \not=0;\ t \in {\bf C}\}$ (with periodicity with respect to the
translations 
$t \to t +2il\pi$). Putting these two facts together, namely the analyticity of
$G(\zeta,t)$ in the union of the sets $S_1$ and $S_2$,  
and making the new change of variables 
$$\alpha= \log {\zeta-\underline \zeta +\epsilon \over \underline \zeta
+\epsilon -\zeta},
\ \  \beta =i \log (t-\underline t),$$ 
one checks that the function $g(\alpha, \beta) \equiv G(\underline \zeta + 
\epsilon {{\rm e}^{\alpha}-1 \over 
{\rm e}^{\alpha}+1},\ \underline t + {\rm e}^{-i\beta})$ is holomorphic in the
following
tube-shaped domain ${\cal T} = {\bf R}^2 +i {\cal B}$, where 
${\cal B}$ is the union of the (disconnected) open set $\{(\Im m\, \alpha, \Im
m\, \beta);\ 
0 < |\Im m\, \alpha| < {\pi\over 2};\ \Im m\, \beta \ {\rm arbitrary}\}$ with   
the ``connection interval''  
$\{(\Im m\, \alpha, \Im m\, \beta);\ 
\Im m\, \alpha =0 ;\ \Im m\, \beta  <\log \rho\}$ (see fig 4).  Now 
since the convex hull of ${\cal B}$ is obviously the domain  
$\check {\cal B} =\{(\Im m\, \alpha, \Im m\, \beta);\ 
{-\pi\over 2} < \Im m\, \alpha < {\pi\over 2};\ \Im m\, \beta \ {\rm
arbitrary}\}$,       
the tube theorem implies that $g(\alpha,\beta)$ admits an analytic continuation 
in ${\bf R}^2 +i \check {\cal B}$, and therefore that $G(\zeta,t)$ admits an
analytic continuation 
in the set 
$ \{ (\zeta,t);\ |\zeta -\underline \zeta| < \epsilon, \  
t \in {\bf C}\}$ (with periodicity with respect to the translations 
$t \to t +2il\pi$). 
Coming back to $F(U,V)$, this shows that $F$ admits an analytic continuation in 
a set which is the union of all complex curves parametrized by  
$U={\rm e}^t,\ V=\zeta {\rm e}^{-t};\ t\in {\bf C}$, for $\zeta$ varying in the 
disk $|\zeta -\underline \zeta| < \epsilon $. These curves are complex
hyperbolae 
except for the one corresponding to the value $\zeta =0$, which is the 
straight-line $V=0$, namely the (complexified) 
``right-mover'' component of the light-cone.  
All these curves can be seen as generated  by the action of all real or complex
Lorentz
transformations (parametrized by $t$) on the set ${\cal N}(\underline p)$   
and Property C is therefore established for the two-dimensional case.

\vskip 1 truecm 
\newdimen\fixhoffset
\fixhoffset=\hsize \relax
\advance\fixhoffset by -16.000\varcm
\divide \fixhoffset by 2
\hbox{\kern\fixhoffset\vbox to 10.00 \varcm{\offinterlineskip
\def\point#1 #2 #3 {\rlap{\kern #1 \varcm
\raise #2 \varcm \hbox{#3}}}
\def\spot{{\kern -0.2em\lower.55ex\hbox{$\bullet$}}}
\vfill\includegraphics{nmicf4.ps}
\smash{\hbox to \hsize{%
\point 13.20 5.00 {$\Im \alpha$}
\point 8.20 9.50 {$\Im \beta$}
\point 8.20 6.40 {$\log \rho$}
\point 8.20 4.50 {0}
\point 4.20 4.50 {$-{\pi \over 2}$}
\point 11.20 4.50 {${\pi \over 2}$}
\hfill}}}\hfill}

\vskip 0.25 truecm
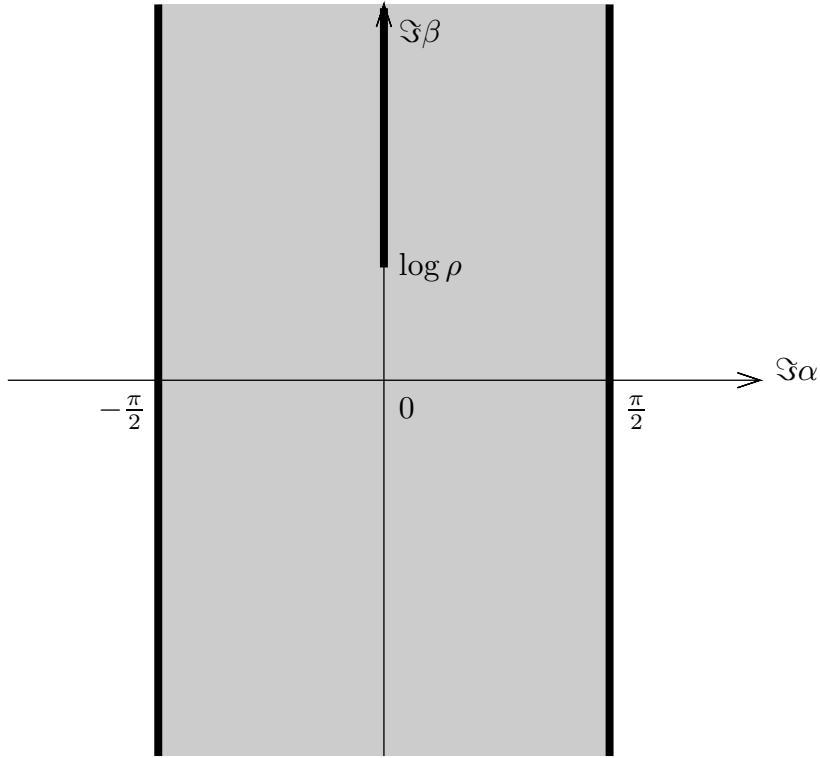
\centerline{Fig.~4. The set ${\cal B}$ (gray)}
\vskip 1 truecm

\vskip 0.1cm
As a by-product of the latter, we stress the following result which is used
below:

\vskip 0.1cm
{\sl Property B with masses:

\noindent
Let ${\cal R}= {\cal R}_{\mu}$ be the set of all 
real energy-momentum vectors $p$ such that $p^2 < {\mu}^2$.  
Then any function $F(k)$ holomorphic in  
$\Sigma_{{\cal R}_{\mu}} =T^+ \cup T^-\cup {\cal R}_{\mu}$ admits an analytic
continuation in 
the domain $\hat \Sigma_{{\cal R}_{\mu}} $ which is the set of all 
{\sl complex} vectors $k$ 
such that $k^2$ belongs to the cut-plane 
${\bf C} \setminus [\mu^2, +\infty[.$}   

{\sl Remark}\ \  It is sufficient that ${\cal R}$ is known to contain
(neighborhoods of) 
one point $\underline p$ on the line $V=0$ and one point 
$\underline p'$ on the line $U=0$ (besides ${\cal R}_0$) 
in order to obtain  
an analyticity domain 
$\hat \Sigma_{\cal R}$ 
of the previous type  
$\hat \Sigma_{{\cal R}_{\mu}} $: in fact, Property C implies that both complex 
lines $U=0$ and $V=0$ are contained in the domain, except maybe for the 
point $U=V=0$ which is not obtained by the previous analytic completion
procedure.
However, this point must also belong to the domain since 
an analytic function of two complex variables {\sl cannot be singular at an
isolated point} surrounded by its domain of analyticity (see e.g. [11]): it
admits an  
analytic continuation at this isolated point defined by an appropriate Cauchy
integral.   

\vskip 0.2cm
{\sl Proof of Properties B and C in the d-dimensional case:}

The general case when $k=(k_0,\vec k)$ varies in ${\bf C}^d$ (e.g. $d=4$ for
field theory in the 
physical Minkowskian space) will be treated by appropriately using the previous 
two-dimensional results  
in sections of ${\bf C}^d$ by (complexified) planes 
containing a time-direction. 

Let $\underline k=\underline p+i\underline q$ be any vector in 
${\bf C}^d$ such that ${\underline k}^2 \in {\bf C} \setminus [0,+\infty[.$ 
In the affine Minkowskian space ${\bf R}^d$ 
consider the point $P$ such that $[OP] \equiv \underline p$ and the 
time-like plane $\Pi$ passing by $P$ and generated by $\underline q$ 
and the unit vector $e_0$ of the time-axis (or choose one of these planes 
and call it $\Pi$ in the degenerate case 
when $\underline q$ is along $e_0$ or is the null vector).  
There is a unique decomposition $\underline p= \underline p' + p_{\perp}$ such
that 
$\underline p'$ is parallel to $\Pi$ and $p_{\perp}$ is orthogonal
to $\Pi$ and therefore spacelike, if not the null vector: 
$p_{\perp}^2 = - \rho^2 \leq 0$. Introducing the 
complexified space $\Pi^{(c)}$ of $\Pi$ and the  
two-dimensional vector variable $k' = p' +iq$ such that every point $k=p+iq$ in 
$\Pi^{(c)}$ can be uniquely written as $ k= k' + p_{\perp}$
with $k'$ orthogonal to $p_{\perp}$, one has:  
${k'}^2= k^2 + \rho^2$.  
In $\Pi^{(c)}$ the section of the domain 
$\Sigma_{{\cal R}_0} =T^+ \cup T^-\cup {\cal R}_0$ is represented in the
vector-variable $k'$ as 
the union of the two-dimensional tubes ${T'}^+$ and ${T'}^-$ defined by ${\Im
m\, k'}^2 >0$ 
and respectively ${\Im m\, k'_0} >0, $ 
${\Im m\, k'_0} <0 $,  and of the real region defined by ${p'}^2 = p^2
-p_{\perp}^2= p^2+ \rho^2
< \rho^2.$ Therefore  
since the given vector 
$\underline k=\underline p' + p_{\perp} +i\underline q \equiv \underline k' +
p_{\perp} $   
is such that ${\underline k'}^2 ={\underline k}^2 + \rho^2 \in {\bf C} \setminus
[\rho^2,+\infty[,$ 
it follows from the two-dimensional {\sl Property B with masses}, applied in  
$k'-$space to the restriction $F'(k') = F_{|\Pi^{(c)}}(k)$ of any function
$F(k)$ analytic in
$\Sigma_{{\cal R}_0}$, that $F'$ admits an analytic continuation at $\underline
k'$ and 
therefore that $F$ itself can be analytically 
continued at the given vector $\underline k$. This shows that Property B 
holds in the $d-$dimensional case. 

Proof of Property C: let us assume that in addition to   
${\cal R}_0$, the set ${\cal R}$ contains 
a given vector $\underline p =[OP]$ with $\underline p^2 \geq 0$. 
Considering at first the case $\underline p^2 >0$, we know that the 
two-sheeted hyperboloid $H(P)$  
with equation $p^2 = \underline p^2$ can be seen as the union of all the
hyperbolae 
$h_{\alpha}(P)$ passing by $P$ 
which are the sections of $H(P)$ by all the two-dimensional planes
$\Pi_{\alpha}$ containing the
parallel to the time axis passing by $P$. In the complexified space of each 
(Minkowskian-type) plane $\Pi_{\alpha}$,
the domain $\Sigma_{\cal R} $ 
admits a restriction represented by a two-dimensional domain of the form
$\Sigma_{{\cal R}_{\alpha}}$, where ${\cal R}_{\alpha}$ contains $P$ 
in addition to a region of the form $p_{\alpha}^2 < \rho_{\alpha}^2$,
corresponding to the intersection of
${\cal R}_0$ by $\Pi_{\alpha}$. Therefore, in view of Property C for the
two-dimensional case 
the whole hyperbola  
$h_{\alpha}(P)$ (and even its complexified) belongs to the holomorphy envelope  
$\hat \Sigma_{{\cal R}_{\alpha}}$ of   
$\Sigma_{{\cal R}_{\alpha}}$.    
Since this is true for all hyperbolae
$h_{\alpha}(P)$, the full hyperboloid $H(P)$ itself belongs to the holomorphy
envelope  
$\hat \Sigma_{\cal R} $ of  
$\Sigma_{\cal R} $.  In the case $\underline p^2 =0$ (with $ P\neq 0)$), $H(P) $
is the light-cone 
and the previous argument 
of analytic completion in the union of all hyperbolic sections by the planes
$\Pi_{\alpha}$ yields 
the whole light-cone {\sl deprived from} the ``light-ray'' distinct from $[OP]$
and  
contained in the (unique) plane $\Pi_{{\alpha}_0}$ passing by the origin.  
However, this exceptional light-ray can be recovered by replacing $P$ by a
neighbouring point $P'$ 
also such that $[OP']^2 =0$: this is always possible since ${\cal R}$ is an open
set.  
(We also note that for the same reason one thus obtains in that case an {\sl
open set} 
$\hat {\cal R}$ of the form $p^2 < \epsilon^2$, the isolated  
point $p=0$ being also obtained 
according to the remark given at the end of the two-dimensional case).   
We have thus 
established the first part of Property C, namely the analytic completion at all
{\sl real} vectors 
$p$ whose corresponding value of $p^2$ is taken by some 
vector $\underline p= [OP] $ 
in ${\cal R}$.  

In order to establish the second part, we can now assume that ${\cal R}$ is the
union of  
${\cal R}_0$ together with a set of 
hypersurfaces $H_{\mu}$ of the form $p^2 = \mu^2,$ with $\mu \geq 0$; 
then there 
remains to prove that all the points of the corresponding {\sl complex} 
hypersurfaces $H_{\mu}^{(c)}$ can be 
reached by the previous analytic completion procedure. Here again, one can
proceed as in the proof of
Property B, namely taking any given vector   
$\underline k=\underline p+i\underline q$ in $H_{\mu}^{(c)}$, one considers the
complex two-dimensional
configuration in the corresponding plane $\Pi^{(c)}$ (specified above in the
proof of Property B).   
Now the section of $H_{\mu}$ by the plane $\Pi$ is a hyperbola contained in the
region ${\cal R}_{\Pi}$ 
of the corresponding section, so that as a result of Property C in the
two-dimensional case, 
the holomorphy envelope contains all the points of the corresponding {\sl
complex} hyperbola,
which includes 
by construction the given point $\underline k$. For 
the case $\mu =0$, the same method still works, including the treatment of the
vectors  
$\underline k=\underline p+i\underline q$ such that $p^2=q^2=0$, which belong to
complexified 
light-rays: the latter are again obtained by the two-dimensional version of
Property C    
in the special case of the right and left movers (no complex light-ray can be
excluded since  
each light-ray has all its real points in the analyticity domain). 
This ends the proof of Property C in the general case.

\vskip 0.5cm
\centerline{\bf 3 Shape of the energy-momentum spectral supports for the
$N-$point functions} 

\vskip 0.3cm
\noindent
We shall now show that the previous study can be repeated  
{\sl for the sector generated by ``two-field vector-states'' of the form 
$\int \varphi(x,x') \Phi(x)\Phi(x')> dxdx'$}. It   
is in fact possible  to perform a similar treatment in complex momentum space, 
in which propagators of the 
fields are now replaced by four-point functions of the latter: the corresponding
results
on the form of dispersion laws will then apply to composite particles appearing
as 
``two-field bound-states''. Subsequently, we shall indicate 
the existence of a similar treatment for the sectors of ``$n-$field 
vector states'' in terms of $2n-$point Green's functions with applications to
dispersion 
laws of composite particles appearing as ``$n-$field bound states'', with $n\geq
3$.  
The validity of such a general study relies in an essential way 
on the general formalism of 
the analytic Green's functions of interacting fields in complex momentum space
[17]. 

The basic fact is that there exists an analog of formula (3) for 
the four-point function, which can be written as follows (see again ${ }^{(3)}$
for our 
use of the bracket notation):
$$ F^+(p; p_1,p_2) - F^-(p; p_1,p_2) = 
<[\tilde R(p_1, p-p_1),  
\tilde R(p_2, -p-p_2)]_{\pm}>\ \ \ \ \eqno(4)$$ 
where $\tilde R$ denotes the Fourier transform of a retarded two-point field
operator 
carrying the total energy-momentum $p$: 
$$\tilde R(p',p-p') = 
\int {\rm e}^{ip\cdot x} 
\ {\rm e}^{ip'\cdot (x'-x)} 
\ \theta(x'_0 -x_0)\ [\Phi(x'), \Phi(x)]_{\pm}\ dx dx' \ \ \ \ \ \eqno(5)$$ 
and where 
$ F^+(p; p_1,p_2)$ and $ F^-(p; p_1,p_2)$ are distributions affiliated with the
{\sl ``generalized 
retarded four-point functions''} (see [8,9]).

\vskip 0.3cm
Here again, 
any usable information on the 
support of the energy-momentum spectrum 
of the theory in the corresponding two-field sector  
will amount to specifying an open subset ${\cal R}$ in the space of 
energy-momentum vectors $(p,p_1,p_2)$ {\sl whose boundary only depend on the
total  
energy-momentum vector} $p$   
in which the distributions  
$<\tilde R(p_1, p-p_1)  
\tilde R(p_2, -p-p_2)>$ 
and 
$<\tilde R(p_2, -p-p_2) 
\tilde R(p_1, p-p_1)>$  
vanish together.  
In view of (4), such a support property (corresponding to the 
knowledge of the ``intermediate states in the latter matrix elements'') then
implies the
{\sl coincidence relation} 
$F^+_{|{\cal R}} = F^-_{|{\cal R}}$.   

\vskip 0.3cm
Moreover, as in the case of propagators, 
the {postulate of microcausality} for the field $\Phi(x)$  
implies {\sl properties of analytic continuation} of the previous  
objects in {\sl complex energy-momentum space}, which play a crucial role.  
These properties hold (together with polynomial bounds)  
in the tempered-distribution setting of Wightman-LSZ fields
considered in [17]; however they could also be justified 
(with a corresponding release of the bounds) in the hyperfunction setting
along the same lines as in Sec. 2.1. 
Even if the description of these properties is more complicated, due to the
occurrence of  
{\sl three complex energy-momenta} 
$k=p+iq,$ $ k_1= p_1 +iq_1,\ k_2=p_2 +iq_2$,  
the situation reproduces 
the case of propagators {\sl as far as the total energy-momentum $p$ is 
concerned}. In fact, $F^+$ and $F^-$ are boundary values of holomorphic
functions from 
tubes ${\cal T}^+,{\cal T}^-$  whose projections onto the space of {\sl complex}
total energy-momentum $k=p+ iq$ are 
respectively $T^+:\ q\in V^+$ and    
$T^-:\ q\in V^-$, so that formula (4) still appears (like (3)) as a 
discontinuity formula: it indicates that 
the discontinuity between the two holomorphic functions $F^+(k;k_1,k_2)$ and 
$F^-(k;k_1,k_2)$
is known to vanish on the set ${\cal R}$.        

\vskip 0.3cm
However we must describe more carefully the situation concerning the analyticity
properties of these functions in the 
{\sl ``internal momenta''}
$k_1$ and $k_2$.   
First, it is clear from formula (5) that in view of the support 
property of the retarded product ($x'-x $ contained in $\overline V^+$), 
$\tilde R$ is the boundary value of an (operator-valued) analytic function  
$\tilde R(k', p-k')$ from the tube $k'=p'+iq':\ q' \in V^+$ for all real $p$.   
Therefore the r.h.s. of Eq.(4) is the 
boundary value of a holomorphic function $\Delta F(p; k_1,k_2)$  of $(k_1,k_2)$ 
in the tube $\Theta$ defined by the conditions $q_1 \in V^+,\ q_2 \in V^+$ for
all real $p$.  

Now it is also shown [8,9] that the domains of analyticity of $F^+,F^-$ implied 
by microcausality are the tubes
${\cal T}^+$ and ${\cal T}^-$ defined by the following conditions:
$${\cal T}^+:\ \ q \in V^+,\  q_1\in V^+,\  q_2 \in V^+\ \ \ \ \ \eqno(6)$$ 
$${\cal T}^-:\ \ -q \in V^+,\  q+q_1\in V^+,\  q+q_2 \in V^+\ \ \ \ \ \eqno(7)$$
and one easily checks that these two tubes admit precisely {\sl as their 
common boundary} (at $q=0$) the tube $\Theta$ for all real $p$. On the latter,
there holds the
following discontinuity formula for the boundary values of $F^+$ and $F^-$:  
$$ \Delta F(p; k_1,k_2) = F^+(p; k_1,k_2) - F^-(p;k_1,k_2).\ \ \eqno(8)$$ 

The main geometrical difference with respect to the case of propagators is that 
the tubes ${\cal T}^+$ and 
${\cal T}^-$ in the big complex $(k,k_1,k_2)-$space {\sl are not opposite} as 
it is the case for $T^+$ and $T^-$ in $k-$space. As a matter of fact, in view of
(6) and (7), the union of  
the tubes ${\cal T}^+$ and 
${\cal T}^+$ admits a convex hull $\check{\cal T}$ which is contained in the 
tube defined by the conditions 
$q_1\in V^+,\  q_2 \in V^+,\  
q+q_1\in V^+,\  q+q_2 \in V^+.$ 
Now in such a situation, and provided  
the {\sl coincidence relation} 
$F^+_{|{\cal R}} = F^-_{|{\cal R}}$ holds true,    
there exists a generalized version of the 
edge-of-the-wedge theorem [10], which states that  
$F^+(k;k_1,k_2)$ and $F^-(k;k_1,k_2)$ still admit a {\sl common analytic
continuation} 
$F(k;k_1,k_2)$. The latter is  
analytic in the union of ${\cal T}^+$, ${\cal T}^-$ and of a complex set ${\cal
N({\cal R})}$ of the
following form: ${\cal N({\cal R})}$ is the intersection of a complex
neighborhood of ${\cal R}$ 
with the convex hull
$\check{\cal T}$ of  
${\cal T}^+ \cup {\cal T}^-$;  
in other words, $F^+$ and $F^-$ ``communicate analytically'' through  
the complex set ${\cal N({\cal R})}$  
which is bordered by ${\cal R}$, although not being analytic anymore in ${\cal
R}$ itself.

\vskip 0.2cm
In the present situation, the open set ${\cal R}$ is always of the following
``cylindric'' form: 
$p_1$ and $p_2$ are arbitrary and $p$ varies in an open set $\underline {\cal
R}$ (namely the
projection of ${\cal R}$ onto $p-$space). Then the equivalence of the following
two statements 
(proved in [8,9]) 
deserves to be stressed:  

a) the boundary values of 
$F^+(k;k_1,k_2)$ and $F^-(k;k_1,k_2)$ 
coincide on ${\cal R}$, 

b) $\Delta F(p;k_1,k_2)$ vanishes {\sl as an analytic function of $(k_1,k_2)$ in
$\Theta$} 
for all $p$ in $\underline {\cal R}$. 

Property b) means that the ``bridge'' in which $F^+$ and $F^-$ have a common
analytic continuation 
contains not only the ``small''set ${\cal N}({\cal R})$ but the ``large common
face'' 
defined by the conditions  $(k_1,k_2)$ in $\Theta$ 
for all $p$ in $\underline {\cal R}$. 

\vskip 0.3cm
As in Sec. 2, one is then led to make use of   
an analytic completion procedure in order to enlarge the 
primitive (``non-natural'') set  
$\Sigma_{\cal R} = {\cal T}^+ \cup {\cal T}^- \cup {\cal N({\cal R})},$  
in which $F(k;k_1,k_2)$ is 
known to be analytic.
It turns out that one can obtain results very similar to those of Sec 2, 
which reproduce the corresponding physical interpretations. 
In fact, the Properties A', B' and C' listed below can be seen as exact
counterparts of the   
respective Properties A, B and C, since they involve identical  
regions (now called) $\underline {\cal R}$ and $\underline {\hat {\cal R}}$ 
in the space of the total energy-momentum $p$, while the
additional analyticity properties with respect to the internal energy-momenta
$k_1$ and $k_2$ 
are a remnant of microcausality in these variables.    

\vskip 0.4cm
i) {\sl Dispersion laws with subluminal velocities}  

\vskip 0.2cm
{\sl Under the} weak {\sl assumption that energy-positivity only holds in} 
privileged {\sl Lorentz frames (see Sec 2-1
and ${ }^{(4)}$),  
microcausality implies that all the hypersurfaces ${\cal M}_i$ and ${\cal M}$
representing  
respectively dispersion laws $p_0=\omega_i(\vec p)$ of one-particle states and
the border of the
continuous energy-momentum spectrum of ``intermediate states in the  
matrix elements'' 
$<\tilde R(p_1, p-p_1)  
\tilde R(p_2, -p-p_2)>$ 
have to be space-like hypersurfaces.} 

This follows from

\vskip 0.4cm
{\sl Property A':  

\noindent
Let ${\cal R}_{a,b}$ be the set of all points $(p,p_1,p_2)$
such that $p$ belongs to a neighborhood of a given  
time-like segment $]a,b[$  
with end-points $p=a$ and $p=b$ 
($b$ in the future of $a$). Then any function $F(k;k_1,k_2)$ holomorphic in  
$\Sigma_{{\cal R}_{a,b}} = {\cal T}^+ \cup {\cal T}^- \cup {{\cal N}({\cal
R}_{a,b})}$  
admits an analytic continuation in a (complex) domain which contains  
the set of all points $(p,k_1,k_2)$ such that $p$ belongs to the double-cone  
$\diamond_a^b$ and $(k_1,k_2)$ varies arbitrarily in the tube $\Theta$.}

\vskip 0.2cm
The argument of Sec 2-1, based on the consideration of time-like segments
$]a,b[$ with $b$ 
contained in ${\cal M}$ or
${\cal M}_1$, then shows again the necessity of the 
space-like character of these hypersurfaces. 
In fact, for all such choices of $]a,b[$, 
the conclusion of Property A' implies that the discontinuity 
$\Delta F(p;k_1,k_2)$ of $F$ vanishes 
for all $p$ in $\diamond_a^b$ and $(k_1,k_2) $ in $\Theta$ and therefore that  
the distribution 
$<[\tilde R(p_1, p-p_1),  
\tilde R(p_2, -p-p_2)]_{\pm}>$ 
vanishes   
for all $p$ in $\diamond_a^b$ and $(p_1,p_2) $ arbitrary.

\vskip 0.4cm
ii) {\sl Lorentz invariance of dispersion laws }  

\vskip 0.2cm
{\sl Under the (usual)} strong {\sl assumption that energy-positivity holds in} 
all {\sl Lorentz frames, 
microcausality implies 
(as in  Sec 2-2)  
that all the hypersurfaces ${\cal M}_i$ and ${\cal M}$ representing  
respectively dispersion laws $p_0=\omega_i(\vec p)$ of one-particle states 
and the border of the continuous energy-momentum spectrum of ``intermediate
states in the  
matrix elements'' 
$<\tilde R(p_1, p-p_1)  
\tilde R(p_2, -p-p_2)>$ 
have to be hyperboloid-shells with equations of the form $p_0= \sqrt {{\vec p}^2
+ m_i^2}$,
$p_0= \sqrt {{\vec p}^2 + M^2}$.}  
This follows from the applicability of  

\vskip 0.2cm
{\sl Property B':   

\noindent
Let ${\cal R}= {\cal R}_0$ be the set of all (real) configurations $(p,p_1,p_2)$
such that the total energy-momentum vector $p= (p_0, \vec p)$ belongs to the
following region       
${\underline {\cal R}}_0:\  |p_0| < |\vec p|$. Then  
any function $F(k;k_1,k_2)$ holomorphic in  
$\Sigma_{{\cal R}_0} ={\cal T}^+ \cup {\cal T}^-\cup {\cal R}_0$ admits an
analytic continuation in 
the domain $\hat \Sigma_{{\cal R}_0} $ which is the set of all 
{\sl complex} configurations $(k,k_1,k_2)$ belonging  
to the convex hull
$\check{\cal T}$ of  
${\cal T}^+ \cup {\cal T}^-$ and   
such that $k^2 \equiv k_0^2 - {\vec k}^2 $ is different from any positive number
and from zero,}   

\vskip 0.2cm
supplemented by

%\eject  
\vskip 0.2cm
{\sl Property C' (Lorentz-invariance of the borders of the spectral supports):  

\noindent 
If $\cal R$ is any real open set, not necessarily connected, containing ${\cal
R}_0$ 
and of ``cylindric form'' $p\in 
\underline {\cal R}$, with  
$\underline {\cal R} \supset   
{\underline {\cal R}}_0$, $p_1,p_2$ arbitrary,  
then 
every function $F(k; k_1,k_2)$ holomorphic in 
$\Sigma_{{\cal R}} ={\cal T}^+ \cup {\cal T}^-\cup {\cal R}$ admits an analytic
continuation in 
the set of all configurations $(p,k_1,k_2)$ such that $(k_1,k_2)$ belongs to the
tube $\Theta$ 
and $p$ varies in an open set 
$\hat {\underline  {\cal R}}$ defined as in Property C: 
it is (for $d\geq 3$) the set of all real vectors $p$ 
whose Minkowskian norm $p^2$ has a value already taken at some vector in 
$\underline {\cal R}$. 
Equivalently (but then including the case $d=2$), it is the set 
of all vectors $p$ obtained from vectors in  
$\underline {\cal R}$ by the action of a (real) Lorentz transformation. }

\vskip 0.2cm
The conclusion of Property C' implies that the discontinuity 
$\Delta F(p;k_1,k_2)$ of the holomorphic function $F(k;k_1,k_2)$ vanishes 
for all $p$ in 
$\hat {\underline  {\cal R}}$  
and $(k_1,k_2) $ in $\Theta$ and therefore that  
the distribution 
$<[\tilde R(p_1, p-p_1),  
\tilde R(p_2, -p-p_2)]_{\pm}>$ 
vanishes   
for all $p$ in  
$\hat {\underline  {\cal R}}$  
and $(p_1,p_2) $ arbitrary.  
It thus expresses the property of Lorentz invariance of the borders of the 
energy-momentum spectrum and therefore (according to the same analysis as in Sec
2-2) 
the results announced above follow.  

\vskip 0.2cm
A derivation of Properties A',B' and C' 
can be given along the same line as the proofs of Properties A,B and C 
presented above in Sec. 2.  
Let us only mention here that Property A' corresponds to a specific case of the 
double-cone theorem for tubes ${\cal T}^+$, ${\cal T}^-$ in general (i.e.
non-opposite) 
situations (see [16])  and that Property B' is exactly the statement given in
Theorem 1 of [8] 
for the case of $n=3$ vector variables, with $m=0$. 

\vskip 0.3cm
{\sl Remark:}\ \  In the statements previously given under i) and ii), the
constraints which 
were obtained concern the shape of the energy-momentum spectrum as it appears in
the subspace of 
two-field states generated by retarded products of the following form  
$R [\varphi]> = 
\int \varphi(x,x') 
\ \theta(x'_0 -x_0)\ [\Phi(x'), \Phi(x)]_{\pm}> \ dx dx'. $ 
\footnote{${ }^{(6)}$}{Rigorously speaking, the passage from support properties
of 
the ``scalar''  distribution\break  
$<\tilde R(p_1, p-p_1)  
\tilde R(p_2, -p-p_2)>$ 
to corresponding support properties of the vector-valued 
distribution $\varphi \to R[\varphi]>$  
relies on a Hilbert-space-norm argument.}     
However, it is clear  
that the same treatment and results are valid as well for two-field 
states generated by the corresponding advanced products, and therefore for 
the subspace generated by all states of the form  
$C [\varphi]> = 
\int \varphi(x,x') 
\ [\Phi(x'), \Phi(x)]_{\pm}> \ dx dx' $ (for all admissible test-functions
$\varphi$).  

\vskip 0.3cm
\noindent
{\sl The general case:}

\vskip 0.2cm
We shall now end this section by explaining  why the previous treatment of 
spectral properties of the space of ``two-field states'' can be generalized to
the 
spaces of ``$n-$field states'' for all $n\geq 3$. Although it is not here the 
right place for presenting this general treatment with all its technical
details, it  
is still possible to indicate briefly how it works.  

The formalism of {\sl generalized retarded operators (g.r.o.)} [17]  
allows one to introduce {\sl generalized absorptive parts}: these are  
expectation values of (anti-) commutators of the following form 
$<[\tilde R_{\alpha}(\{p_i;\ i\in I\}),  
\tilde R_{\alpha'}(\{p'_{i'};\ i'\in I'\})]_{\pm}>$,   
where the operators 
$\tilde R_{\alpha}(\{p_i;\ i\in I\})$ and 
$\tilde R_{\alpha'}(\{p'_{i'};\ i'\in I'\})$   
denote the Fourier transforms of $n-$point g.r.o. 
$R_{\alpha}$, $R_{\alpha'}$ with supports  
contained in relevant corresponding salient cones 
${\cal C}_{\alpha}$
and ${\cal C}_{\alpha'}$ in the space of differences $x_j-x_k$ 
(resp. $x'_{j'} - x'_{k'}$) of space-time vectors:  
these cones are (non-trivial) analogs of the supports of the usual
retarded and advanced operators of the case $n=2$ (i.e. $x_1 -x_2 \in {\bar
V}^{\pm}$).  
In our notation, $I$ and $I'$ represent disjoint subsets of $n$ elements
($|I|=|I'|=n$)
of the set $\{1,2,,\cdots, 2n\}$ and the corresponding energy-momenta $p_i,
p'_{i'}$ are 
linked by the energy-momentum conservation law 
$p=\sum_{i\in I} p_i
=-\sum_{i'\in I'} p'_{i'}$, $p$ being the  
total energy-momentum of the corresponding channel $(I,I')$ of the 
$2n-$point function of the fields considered; as previously (see ${ }^{(3)}$), 
it is understood that the distribution  
$\delta(\{\sum_{i\in I} p_i\} + \{\sum_{{i'}\in I'} p'_{i'}\})$    
has been factored out 
in the brackets $<\  >$.  

\vskip 0.3cm 
We then claim that for each $n$ and each $(I,I')$ 
there exists a complete set of g.r.o.  
$R_{\alpha}$, $R_{\alpha'}$ whose Fourier transforms   
satisfy a discontinuity formula analogous to (4) of the following form  
$$ F^+_{\alpha,\alpha'}(\{p_i;\ i\in I\}; \ \{p'_{i'};\ i'\in I'\})  
-F^-_{\alpha,\alpha'}(\{p_i;\ i\in I\}; \ \{p'_{i'};\ i'\in I'\}) = $$ 
$$<[\tilde R_{\alpha}(\{p_i;\ i\in I\}),  
\tilde R_{\alpha'}(\{p'_{i'};\ i'\in I'\})]_{\pm}>;\ \ \ \ \ \eqno(9)$$ in the
latter,   
$ F^{\pm}_{\alpha,\alpha'}(\{p_i;\ i\in I\}; \ \{p'_{i'};\ i'\in I'\}) $   
are distributions affiliated with the {\sl ``generalized 
retarded $2n-$point functions''}   
which are boundary values of analytic functions (still denoted by)  
$ F^{\pm}_{\alpha,\alpha'}(\{k_i;\ i\in I\}; \ \{k'_{i'};\ i'\in I'\}) $   
from respective tubes
${\cal T}^+_{\alpha,\alpha'},$  
${\cal T}^-_{\alpha,\alpha'},$ 
in the space of complex vectors $k_i= p_i+ iq_i, \ k'_{i'}= p'_{i'} +i q'_{i'},$
such that
$k =p+iq=\sum_{i\in I} k_i
=-\sum_{i'\in I'} k'_{i'}.$ 
These pairs of tubes 
play the same role as the pair 
(${\cal T}^+,$  
${\cal T}^-$) of the case of two-field states: all points in   
${\cal T}^+_{\alpha,\alpha'}$ 
(resp. ${\cal T}^-_{\alpha,\alpha'},$)  
satisfy the condition $q= \Im m k \in V^+$ (resp. $V^-$).  
Microcausality is a basic ingredient in the proof of the previous statement,
which relies on 
the results of [17 d), e)].  

\vskip 0.3cm
We are again led to express  
the energy-momentum spectral assumptions   
of the theory in the corresponding $n-$field sector  
by specifying an open subset ${\cal R}$ in the space of 
energy-momentum vectors $p_i,\ p'_{i'} $ whose boundary {\sl only depend on the
total  
energy-momentum vector} $p =\sum _{i\in I} p_i,$   
in which the distributions  
$<\tilde R_{\alpha}(\{p_i;\ i\in I\})  
\tilde R_{\alpha'}(\{p'_{i'};\ i'\in I'\})>$ and   
$<\tilde R_{\alpha'}(\{p'_{i'};\ i'\in I'\})  
\tilde R_{\alpha}(\{p_{i};\ i\in I\})>$   
vanish together. 
Here again, the edge-of-the-wedge theorem [10] implies that  
$ F^+_{\alpha,\alpha'}$  
and $ F^-_{\alpha,\alpha'}$ have a common analytic continuation 
$ F_{\alpha,\alpha'}$  
in a set of the form 
$\Sigma_{\cal R} = {\cal T}^+_{\alpha,\alpha'}   
\cup {\cal T}^-_{\alpha,\alpha'} \cup {\cal N}({\cal R}).$

\vskip 0.3cm
One could then present the ``$n-$field-state version'' of Properties A', B' and
C' 
in a way which closely parallels the two-field state case. For brevity , we
shall not repeat 
the full statements and the corresponding physical interpretations which are
identical to 
those listed above in paragraphs i) and ii) under the respective ``weak'' and
``strong''
forms of the energy-positivity condition. To exhibit the parallelism of the
geometry of the   
$n-$field case with the one of the two-field case, it is sufficient to make a
little more precise 
the description of the situation in the sets of energy-momentum vectors $k_i$
and $k'_{i'}$
and the characterization of the domains  
${\cal T}^+_{\alpha,\alpha'},$  
${\cal T}^-_{\alpha,\alpha'},$ and of their common face in the subspace $k=p$
real.  

For $p$ real, we introduce 
the sets of complex vectors 
$K_I= \{\underline k_i =k_i- {p\over n};\ i\in I\}$ and  
$K'_{I'}= \{\underline k'_{i'} =k'_{i'} +{p\over n};\ i'\in I'\}$   
linked by the relations 
$\sum_{i\in I}\underline k_i =  
\sum_{i'\in I'} \underline k'_{i'} = 0$; correspondingly 
$Q_I= \Im m K_I$ 
(resp.$Q'_{I'}= \Im m K'_{I'}$) is the set of all $q_i$ (resp. $q'_{i'}$) such
that 
$\sum_{i\in I} q_i =0$  
(resp. $\sum_{i'\in I'} q'_{i'} = 0$). 
Each of the sets of vectors $K_I$, $K'_{I'}$ (resp.  
$Q_I$, $Q'_{I'}$) varies in a space of $(n-1)$ independent complex (resp. real) 
energy-momentum vectors.

By taking into account analogs of formula (5) 
for the operators  $\tilde R_{\alpha}$ 
and $\tilde R_{\alpha'}$ 
together with linear identities between 
them (called ``Steinmann relations'' [17]), one can deduce from the support
properties  
of $R_{\alpha}$ and
$R_{\alpha'}$  
(namely supp $R_{\alpha} \subset {\cal C}_{\alpha}$,  
supp $R_{\alpha'} \subset {\cal C}_{\alpha'}$) the following analyticity
property: 
the r.h.s. of Eq.(9) is for every real $p$ the boundary value of an analytic 
function $\Delta F_{\alpha,\alpha'}(p; K_I,K'_{I'})$ of $(K_I,K'_{I'}),$  
holomorphic in a well-defined tube $\Theta_{\alpha,\alpha'}$ (playing the same
role as $\Theta$ in
the case $n=2$). This tube is specified by 
a set of conditions of the  following type in the space of the 
imaginary parts $(Q_I,Q_{I'})$. There exists 
a set $\Pi_{\alpha}$ of partitions $(J,L)$ of $I$ and  
a set $\Pi'_{\alpha'}$ of partitions $(J',L')$ of $I'$ such that
the defining conditions for 
$\Theta_{\alpha,\alpha'}$ are:  
$q_J =-q_L \in V^+$ and $q'_{J'}= -q'_{L'} \in V^+$   
for all $(J,L)$ in $\Pi_{\alpha}$ and   
all $(J',L')$ in $\Pi'_{\alpha'}$: in the latter the notation 
$q_J$ (resp. $q'_{J'}$) refers to the corresponding partial sum     
$\sum_{i\in J} q_i$ (resp.  
$\sum_{i'\in J'} q'_{i'}$).  The sets  
$\Pi_{\alpha}$ and  
$\Pi'_{\alpha'}$ are not arbitrary but must satisfy 
the so-called ``cell-conditions'' (see [17]) which 
express the fact that no linear subspace with equation $q_{M}=0 $ 
or $q'_{M'}=0$, with $M\subset I$ and  
$M'\subset I'$ intersects the domain     
$\Theta_{\alpha,\alpha'}$.  

\vskip 0.2cm
Now it can be shown that the 
tubes ${\cal T}^+_{\alpha,\alpha'}$  
and ${\cal T}^-_{\alpha,\alpha'}$ in which the functions  
$ F^+_{\alpha,\alpha'}$  
and $ F^-_{\alpha,\alpha'}$ are holomorphic are defined by the 
following conditions:
$${\cal T}^+_{\alpha,\alpha'}:\ \ q\in V^+,\    
-q_L \in V^+\  {\rm and}\  q'_{J'} \in V^+ \ \ \ \eqno(10)$$   
for all $(J,L)$ in $\Pi_{\alpha}$ and   
all $(J',L')$ in $\Pi'_{\alpha'};$
$${\cal T}^-_{\alpha,\alpha'}:\ \ -q\in V^+,\    
q_J =-q_L +q \in V^+\  {\rm and}\  q'_{J'}+q =-q'_{L'} \in V^+\ \ \eqno(11)$$   
for all $(J,L)$ in $\Pi_{\alpha}$ and   
all $(J',L')$ in $\Pi'_{\alpha'}.$ 

\vskip 0.2cm
These two tubes admit as their 
common boundary (at $q=0$) the tube $\Theta_{\alpha,\alpha'}$ for all real $p$. 
On the latter, there holds the
following discontinuity formula for the boundary values of 
$ F^+_{\alpha,\alpha'}$  
and $ F^-_{\alpha,\alpha'}$:  
$$\Delta F_{\alpha,\alpha'}(p; K_I,K'_{I'})= $$  
$$F^+_{\alpha,\alpha'}(\{k_i;\ i\in I\}; \ \{k'_{i'};\ i'\in I'\})_{|q=0}    
- F^-_{\alpha,\alpha'}(\{k_i;\ i\in I\}; \ \{k'_{i'};\ i'\in I'\})_{|q=0}.\ \ \
\eqno(12)$$  

One easily checks that the defining conditions (10), (11) 
of the tubes ${\cal T}^+_{\alpha,\alpha'}$  
and ${\cal T}^-_{\alpha,\alpha'}$   
are completely analogous to 
the defining conditions (6), (7) of ${\cal T}^+$ and  ${\cal T}^-$, up to the 
replacement of the two vector variables $q_1,\ q_2$  by all the vector variables
$-q_L,\ q'_{J'}$  corresponding to the sets of partitions $\Pi_{\alpha}$,
$\Pi_{\alpha'}$.  

As a matter of fact, it is known (see [17]) that it is sufficient to consider a 
subset of g.r.o. called ``Steinmann monomials'' 
$R_{\alpha}$, $R_{\alpha'}$  for which each of the corresponding sets    
$\Pi_{\alpha}$, $\Pi_{\alpha'}$ contains {\sl exactly} $n-1$ partitions (one
also 
says that the corresponding cell-conditions are ``simplicial'');  
in fact, the most general g.r.o. are linear combinations of these Steinmann
monomials.  
It then turns out that in this restricted class of g.r.o. the analog of 
Property B' 
coincides with Theorem 1 of [8] in its general $n-$vector form (with $m=0$): 
this property states that {\sl any function holomorphic in 
$\Sigma_{{\cal R}_0} = {\cal T}^+_{\alpha,\alpha'}   
\cup {\cal T}^-_{\alpha,\alpha'} \cup {\cal N}({{\cal R}_0})$ (with 
${{\cal R}_0} $ now defined by the conditions $|p_0| < |\vec p|$, 
$K_I$ and $K'_{I'}$ real and arbitrary), admits an analytic continuation at  
all the points $(k, K_I, K'_{I'})$ in the convex hull of the tube  
${\cal T}^+_{\alpha,\alpha'}   
\cup {\cal T}^-_{\alpha,\alpha'}$  
such that $k^2\equiv k_0^2 - \vec k^2$ is different from any positive number and
from zero.} 
Property C' then follows from B' as in the case $n=2$, 
while Property A' corresponds again to the double-cone theorem in a  
geometrical situation of general type.  

These considerations can be completed by a remark similar to the one 
given at the end of the case $n=2$ (including footnote ${ }^{(6)}$): 
since the g.r.o. generate (by linear 
combinations of Steinmann monomials) all the multiple (anti-)commutators of 
$n$ field operators, the constraints  on the energy-momentum spectrum 
apply to the subspace generated by all states of the form    
$C[\varphi]> = \int \varphi(x_1,\ldots, x_{n-1},x_n) 
[\Phi(x_1),[\ldots,[\Phi(x_{n-1},\Phi(x_n)]..]] > \ dx_1\ldots dx_n. $  (for all
admissible test-functions $\varphi$).

\vskip 0.3cm

\centerline{\bf 4 Concluding remarks}  
\vskip 0.2cm
In this paper, we have displayed the geometrical constraints on the 
shape of the energy-momentum
spectrum which 
result from microcausality together with (weak or strong)
energy-positivity requirements in any (boson or fermion) interacting field
theory. The Lorentz-invariant shape of the spectrum is therefore proven 
for possibly existing cases of field theories with Lorentz symmetry breaking 
with a high degree of generality.
This is due to the purely geometrical character of our method, 
based on analyticity properties in 
several complex variables, which has allowed a strict exploitation of the latter
requirements {\sl in terms of the analytic  
Green's functions of the fields}: it is in terms of these objects that 
the spectral constraints are expressed and the degree of generality which is reached 
in this approach presents various aspects. 

As already noticed 
(see our Remark in  Sec. 3 and footnote ${ }^{(6)}$),  
the Hilbert space interpretation 
of these constraints can be done separately.  
An advantage of the  
Green's function approach is
therefore the fact that the constraints obtained are still proven to hold 
in an indefinite-metric framework, as for example in the usual treatment of the 
QCD-fields with a gauge-fixing preserving the microcausality conditions for the 
Green's functions.  

Concerning the fact, already stressed above, that the results apply  
to the case of possible nonconventional field theories 
with arbitrarily wild short-distance singularities (i.e. of hyperfunction type, as
specified in Sec. 2.1), we can say that the generality of the results
goes even further: the latter are valid as soon as  
the Green's functions are analytic in the relevant 
tube domains of momentum space without any restriction on the increase
of these functions at infinity, which can be considered as the most general 
expression of microcausality (note that mild exponential bounds in the energy 
were still obtained in the $x-$space hyperfunction setting, 
as noticed at the end of Sec. 2.1).  
In this connection,  
the method applies for example   
to the  special case
of Green's functions 
enjoying a very regular behaviour at large real energy-momenta,  
but exponentially increasing at purely imaginary energies: this is precisely 
the case for the Green's functions  
of the fields generated (via space-time translations) by local observables in
the  
``local quantum physics'' framework of [4], which were considered in the original
works by 
Borchers and Buchholz [2,3] on the present subject.

\vskip 0.6cm
\centerline{\bf References}

\vskip 0.6cm
\noindent
[1] V.A. Kosteleck\'y and R. Lehnert, Phys. Rev. D {\bf 63}, 065008   
(2001).  

\vskip 0.2cm
\noindent
[2] H.J. Borchers and D. Buchholz, Commun. Math. Phys. {\bf 97}, 169-185 (1985).

\vskip 0.2cm
\noindent
[3] H.J. Borchers, Fizika (Zagreb) {\bf 17}, 289-304 (1985).

\vskip 0.2cm
\noindent
[4] R. Haag, {\sl Local Quantum Physics}  (Springer-Verlag, 
Berlin-Heidelberg-New York, 1992).

\vskip 0.2cm
\noindent
[5] H.J. Borchers, Commun. Math. Phys. {\bf 22}, 49-54 (1962).

\vskip 0.2cm
\noindent
[6] H.J. Borchers, {\sl Translation Group and Particle Representations in
Quantum Field\break  
\hskip 1cm
\quad Theory }  
(Springer, New York, 1996).

\vskip 0.2cm
\noindent
[7] R. Jost and H. Lehmann, Nouvo Cimento {\bf 5}, 1598-1610 (1957); F.J. Dyson,
Phys. Rev. 
\hskip 1cm
{\bf 
\quad 110}, 1460-1464 (1958).

\vskip 0.2cm
\noindent
[8] J. Bros , H. Epstein and V. Glaser, Nuovo Cimento  {\bf 31 }, 1265-1302
(1964).

\vskip 0.2cm
\noindent
[9] H. Epstein, {\sl Some analytic properties of scattering amplitudes in
quantum field theory} 
in ``Particle Symmetries and Axiomatic Field Theory'', Brandeis Summer School
1965,(Gordon and Breach,   
New York,1966). 

\vskip 0.2cm
\noindent
[10] H. Epstein, Journ. Math. Phys. {\bf 1}, 524-531 (1960).   

\vskip 0.2cm
\noindent
[11] S. Bochner and W.T. Martin, {\sl Several complex variables} (Princeton
University Press, Princeton, NJ, 1948). 

\vskip 0.2cm
\noindent
[12] A.S. Wightman, {\sl Analytic functions of several complex variables} 
in ``Relations de dispersion et  
particules \'el\'ementaires'', Les Houches Summer School 1960, C. De Witt and R.
Omnes eds 
(Hermann, Paris, 1960).

\vskip 0.2cm
\noindent
[13] L. Asgeirsson, Math. Ann. {\bf 113}, 321 (1936).

\vskip 0.2cm
\noindent
[14] V.S. Vladimirov, Doklady Akad. Nauk SSSR {\bf 134}, 251 (1960).

\vskip 0.2cm
\noindent
[15] H.J. Borchers, Nuovo Cimento {\bf 19}, 787-796 (1961).

\vskip 0.2cm
\noindent
[16] J. Bros , H. Epstein, V. Glaser and R.Stora, in ``Hyperfunctions and 
Theoretical Physics'',  Lecture Notes in Mathematics {\bf 449}, 185 
(Springer-Verlag, Berlin, 1975).

\vskip 0.2cm
\noindent
[17]\ -a) D. Ruelle, Nuovo Cimento {\bf 19}, 356 (1961) and Thesis, Zurich 1959;

-b) O. Steinmann, Helv. Phys. Acta {\bf 33}, 257 (1960); {\bf 33}, 347 (1960); 

-c) H.Araki and N. Burgoyne, Nuovo Cimento {\bf 18}, 342 (1960); H. Araki, J.
Math. Phys. 
{\bf 2}, 163 (1961); 

-d) J. Bros, Comptes-rendus RCP no 25, CNRS Strasbourg (1967) and Thesis, Paris
1970; 

-e) H. Epstein, V. Glaser and R. Stora, {\sl General Properties of the n-point
Functions in  
Local Quantum Field Theory} in ``Structural Analysis of Collision Amplitudes'',
Les Houches 
June Institut 1975, R. Balian and D. Iagolnitzer eds (North-Holland, Amsterdam,
1976).

\vskip 0.2cm
\noindent
[18] R. F. Streater and A. S. Wightman, {\sl PCT, Spin and Statistics and all that},    
Benjamin,  New York (1964) 

\vskip 0.2cm
\noindent
[19] H. Lehmann, K. Symanzik and W. Zimmermann, Nuovo Cimento {\bf 1}, 205 (1955) and
{\bf 6}, 319  (1957)

\vskip 0.2cm
\noindent
[20] S. Nagamachi and N. Mugibayashi, Letters in Math. Phys. {\bf 1}, 259-264 (1976).

\vskip 0.2cm
\noindent
[21] A. Jaffe,  Phys. Rev. {\bf 158}, 1454 (1967).

\end